\documentclass[11pt,preprint]{aastex}

\begin{document}

\shorttitle{V4046 Sgr Disk Structure}
\shortauthors{Rosenfeld et al.}

\title{The Structure of the Evolved Circumbinary Disk around V4046 Sgr}

\author{
Katherine A. Rosenfeld\altaffilmark{1},
Sean M. Andrews\altaffilmark{1},
David J. Wilner\altaffilmark{1},
J. H. Kastner\altaffilmark{2}, \&
M. K. McClure\altaffilmark{3}
}
\altaffiltext{1}{Harvard-Smithsonian Center for Astrophysics, 60 Garden Street, Cambridge, MA 02138}
\altaffiltext{2}{Center for Imaging Science, Rochester Institute of Technology, 54 Lomb Memorial Drive, Rochester, NY 14623}
\altaffiltext{3}{Department of Astronomy, University of Michigan, 830 Dennison Bldg, 500 Church Street, Ann Arbor, MI 48109}

\begin{abstract}
We present sensitive, sub-arcsecond resolution Submillimeter Array observations 
of the protoplanetary disk around the nearby, pre-main sequence spectroscopic 
binary V4046 Sgr.  We report for the first time a large inner hole ($r=29$\,AU) 
spatially resolved in the 1.3\,mm continuum emission and study the structure of 
this disk using radiative transfer calculations to model the spectral energy
distribution (SED), continuum visibilities, and spectral line emission of CO and
its main isotopologues.  Our modeling scheme demonstrates that the majority of 
the dust mass is distributed in a narrow ring (centered at 37\,AU with a FWHM of
16\,AU) that is $\sim$5$\times$ more compact than the gas disk.  This structure 
implies that the dust-to-gas mass ratio has a strong spatial variation, ranging 
from a value much larger than typical of the interstellar medium (ISM) at the 
ring to much smaller than that of the ISM at larger disk radii.  We suggest that
these basic structural features are potentially observational signatures of the 
accumulation of solids at a local gas pressure maximum.  These models also 
require a substantial population of $\sim$$\mu$m-sized grains inside the central
disk cavity.  We suggest that this structure is likely the result of dynamical 
interactions with a low-mass companion, although photoevaporation may also play 
a secondary role.
\end{abstract}
\keywords{circumstellar matter --- protoplanetary disks --- planet-disk 
interactions --- submillimeter: planetary systems 
--- stars: individual (V4046 Sgr)}

\section{Introduction} \label{sec:intro}

In the past few years, high angular resolution millimeter/radio-wave
observations have facilitated rapid development in our understanding of some
fundamental aspects of protoplanetary disk evolution.  For example, there is
mounting evidence for substantial discrepancies between the spatial
distributions of mm-sized dust particles and molecular gas in protoplanetary
disks, possibly caused by a dramatic decrease in the dust-to-gas ratio at large 
radii \citep[e.g.,][]{panic09,andrews12}.  Coupling those results with new 
analyses that identify a systematic decrease in dust particle sizes in the 
outer disk \citep{guilloteau11,perez12}, there seems to be good progress toward 
the promise of empirical constraints on the key processes tied to the growth 
and migration of disk solids.  Meanwhile, there has been marked improvement in 
the characterization of another key factor in disk evolution: dynamical 
interactions with companions.  \citet{harris12} provided a sweeping 
quantitative confirmation of the seminal work by \citet{jensen96}, 
demonstrating that tidal interactions in multiple star systems are a prominent 
issue for disk survival, especially for separations in the $\sim$3-30\,AU 
range.  And \citet{andrews11} (among others) have speculated that the ring-like 
millimeter-wave dust continuum emission morphologies noted for the so-called 
``transition" disks are most likely created by an analogous dynamical 
interaction process, though in this case the perturbers may be embedded {\it 
planetary} companions.

For the time being, most of these data-driven studies of disk evolution are 
somewhat piecemeal in terms of both diagnostics and target samples, due to 
practical observational limitations.  While that should soon change 
dramatically with the completion of the Atacama Large Millimeter Array (ALMA)
facility, it is important to recognize that some individual targets will always
serve as particularly illuminating case studies for different aspects of disk
evolution.  Here, we focus on the disk around V4046 Sgr, a remarkable system
with observable characteristics that are likely simultaneously shaped by {\it 
all} of the fundamental processes involved in disk evolution.

V4046 Sgr is a double-lined spectroscopic binary with a 2.4-day orbital period 
\citep{byrne86,delareza86,quast00,stempels04}.  Radial velocity monitoring
indicates a nearly equal-mass pair of K-type stars on a close ($a \approx 
0.045$\,AU), circular ($e \le 0.001$) orbit \citep{stempels13}.
\citet{kastner11} suggested that GSC 07396$-$00759 could be an additional
distant ($\sim$2\farcm8) companion, which itself might be an unresolved close
binary \citep[see also][]{nataf10}.  Two independent, dynamical methods to
estimate the V4046 Sgr stellar masses are in agreement, with $M_{\ast,1} = 
0.90\pm0.05$ and $M_{\ast,2} = 0.85\pm0.04$\,M$_{\odot}$ for the primary and
secondary, respectively \citep{rosenfeld12,stempels13}.  Coupling these
dynamical mass measurements to the inferred temperatures and luminosities,
pre-main sequence evolution models suggest that V4046 Sgr has an age of
$\sim$10-20\,Myr \citep{rodriguez10,donati11,rosenfeld12}.  Those ages are 
consistent with the conjecture of \citet{torres06,torres08} that V4046 Sgr is a 
member of the $\beta$ Pic moving group, with a kinematic parallax distance of 
only 73\,pc from the Sun.  Evidence for a disk around the V4046 Sgr binary first
came from emission line accretion signatures \citep[H$\alpha$ equivalent widths 
of $\sim$30-120\AA;][]{merrill50,henize76,hbc88} and an infrared continuum 
excess in the {\it IRAS} bands \citep[e.g.,][]{johnson86,weintraub90,weaver92}. 
Strong (sub)millimeter emission was detected by \citet{jensen96}, suggesting a 
relatively large disk mass.  \citet{jensen97} modeled the infrared SED and 
concluded that it was consistent with an extended circumbinary disk truncated at
an inner edge radius of $\sim$0.2\,AU, as would be expected from dynamical 
interactions with the central binary \citep[e.g.,][]{artymowicz94}.  Later, 
\citet{kastner08} discovered a substantial reservoir of molecular gas in the 
disk orbiting V4046 Sgr, which was subsequently imaged and found to span 
$\sim$800\,AU in diameter 
\citep[roughly 10\arcsec\ on the sky;][]{rodriguez10,oberg11,rosenfeld12}.  
Fitting an elliptical Gaussian to their 230\,GHz continuum visibilities, 
\citet{rodriguez10} suggested that the mm-wave dust emission 
is concentrated inside a $\sim$40\,AU radius.

Taken together, these properties make the V4046 Sgr system a significant 
benchmark for studies of protoplanetary disk evolution for three
key reasons.  First, the advanced age of the disk suggests that evolutionary
effects have had more time to progress, and therefore should exhibit more
obvious observational signatures than for a typical $\sim$1\,Myr-old disk.
Evolutionary mechanisms like particle growth and migration
\citep[e.g.,][]{birnstiel12a}, giant planet formation
\citep[e.g.,][]{pollack96,hubickyj05}, and gas dissipation via photoevaporative
winds \citep[e.g.,][]{clarke01,alexander09} should all be relevant in shaping
disk properties at the V4046 Sgr age.  For reference, such disks are rare: V4046
Sgr harbors the {\it only} gas-rich disk in the $\beta$ Pic moving group, and 
is one of a handful of gas-rich disks known to be associated with
T Tauri stars in local young stellar groups \citep[the others orbit TW Hya, MP 
Mus, and T Cha;][]{kastner97,kastner10,sacco13}.
Second, the proximity of 
V4046 Sgr is a substantial practical advantage in measuring key evolutionary 
diagnostics: the same observations would be $4\times$ more sensitive and probe 
2$\times$ smaller spatial scales for V4046 Sgr compared to disks around 
younger T Tauri stars that are associated with the nearest star-forming 
clouds (at distances $\sim 140$\,pc).  And third, from the perspective of a 
stellar host, V4046 Sgr introduces some interesting environmental issues that 
could influence the evolution of its disk.  Aside from dynamical clearing due to
its stellar multiplicity, the binary at the center of the V4046 Sgr disk makes 
for an unusual dichotomy.  The combined mass of the two central stars means that
dynamical timescales in the disk are relatively short.  However, the 
irradiation environment is not much different than for a single star, so 
timescales tied to thermal, energetic, or chemical processes are comparatively 
unaffected.  In essence, the evolutionary behavior of the V4046 Sgr disk can be 
described as a hybrid of a typical Herbig Ae disk and T Tauri disk, depending on
the relevant timescale that dominates a given evolution mechanism.

In this article, we present new, sensitive, high angular resolution observations
of the 1.3\,mm continuum and $^{12}$CO/$^{13}$CO/C$^{18}$O $J$=2$-$1 line 
emission from the V4046 Sgr circumbinary disk.  Using these data and a suite of 
radiative transfer tools, we aim to construct a preliminary, representative 
model of the disk structure in an effort to help characterize the observational 
signatures of different disk evolution mechanisms.  In the following sections, 
we describe our observations with the Submillimeter Array (SMA) and the relevant
data calibration procedures (\S 2), present some basic observational results 
(\S 3), develop models for the disk structure (\S 4), and comment on their 
implications for our understanding of disk evolution (\S 5).  A summary is 
provided in \S 6.

\section{Observations and Data Reduction} \label{sec:observations}

We observed V4046 Sgr with the Submillimeter Array \citep[SMA;][]{ho04} on
Mauna Kea, Hawaii on four occasions in 2009 and 2011.  These observations and
their calibration were already described by \citet{rodriguez10} and
\citet{rosenfeld12}, but a brief summary of the key points is provided here for
completeness.  In these observations, the individual 6-m array elements were
arranged in each of the four available SMA configurations, spanning baseline 
lengths from 8 to 509\,m.  The dual-sideband receiver backends and SMA 
correlator were configured with a local oscillator (LO) frequency of 
225.360\,GHz (1.33\,mm) and a $\sim$2\,GHz-wide intermediate frequency (IF) 
band $\pm$4-6\,GHz from the LO: in 2011, a second IF band was included 
$\pm$6-8\,GHz from the LO.  Each sideband/IF band combination was composed of 
24 spectral chunks of 104\,MHz width (although typically only the central 
82\,MHz are used).  In the first IF band, three chunks (in each sideband) were 
split into 512 spectral channels, to sample the CO isotopologue emission at 
200\,kHz ($\sim$0.25\,km s$^{-1}$) resolution.  All other chunks were coarsely 
split into 32 channels (3.25\,MHz each) to observe the continuum.  Observations 
of V4046 Sgr were interleaved with visits to J1924-292 ($\sim$15\degr\ away) on 
a 5-15\,minute cycle, as well as J1733-130 (22\degr\ away) on a 
$\sim$45\,minute cycle.  Additional observations of 3C 454.3 and available 
planets/satellites (Neptune, Ceres, Callisto) were conducted for calibration 
purposes.  A summary of relevant observational parameters is provided in Table 
\ref{tab:obs}.

The visibility data in each IF band and from each SMA observation were
calibrated independently using the {\tt MIR} package, as described by
\citet{rosenfeld12}.  After confirming their consistency over all IF/sideband
combinations and on overlapping baseline lengths, the continuum data were
spectrally averaged.  Spectral visibilities that cover each CO isotopologue
transition were continuum-subtracted and combined.  The overall data quality is
exceptional for the SMA and the low declination of V4046 Sgr, due primarily to 
the excellent observing conditions (precipitable water vapor levels were only 
$\sim$1\,mm throughout all of the observations).  Synthesis images were made 
using the {\tt CLEAN} deconvolution algorithm in the {\tt MIRIAD} software 
package for each emission tracer.  For the 1.3\,mm continuum, we emphasized the 
emission on smaller spatial scales with a Briggs robust (-1) weighting scheme, resulting in a map with a
$0\farcs74\times0\farcs38$ synthesized beam (at P.A. = 14\degr) and an RMS
noise level of 1.0\,mJy beam$^{-1}$.  Channel maps of the $^{12}$CO line 
emission were generated with natural weighting to produce a 
$1\farcs1\times0\farcs9$ beam with an RMS noise level of 40\,mJy beam$^{-1}$ in
25 binned 0.4\,km s$^{-1}$ velocity channels, centered on the systemic LSR
velocity, $+$2.87\,km s$^{-1}$ \citep{rodriguez10}.  Analogous channel maps of 
the CO isotopologue lines were made by employing a 0\farcs5 Gaussian taper, 
producing a slightly larger synthesized beam ($1\farcs5\times1\farcs2$) and a 
similar RMS noise level ($\sim$35\,mJy beam$^{-1}$).

\section{Results} \label{sec:results}

\begin{figure}[t!]
\epsscale{0.95}
\plotone{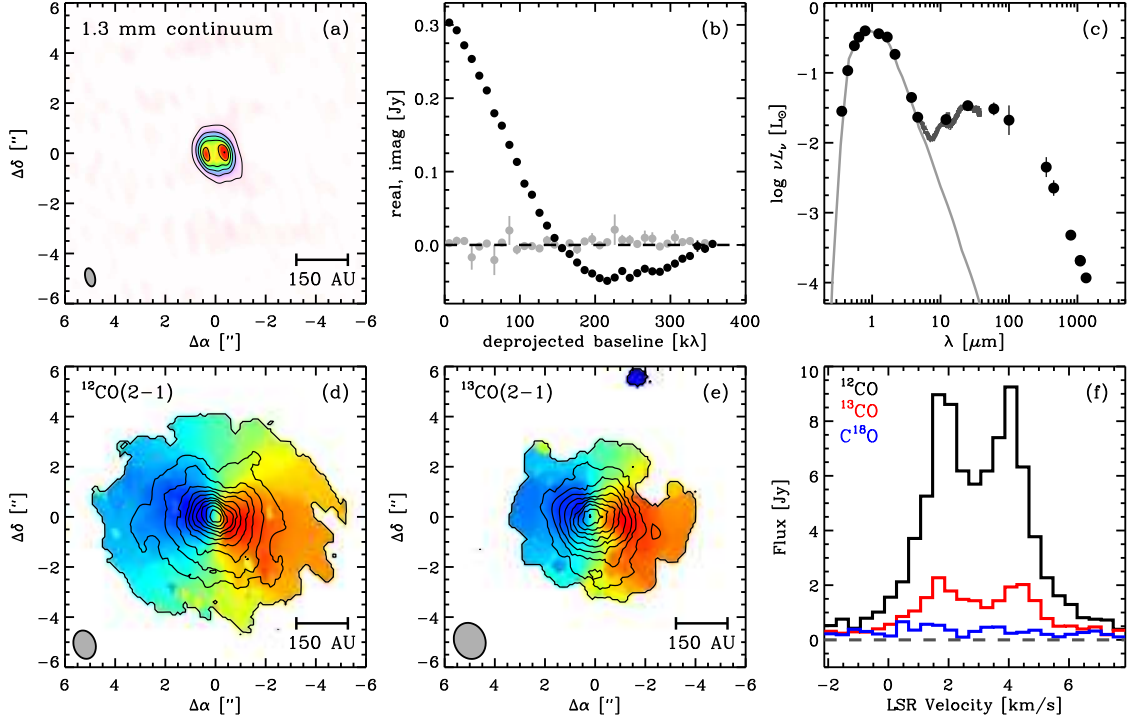} \figcaption{An observational summary of the V4046
Sgr disk.  (a) An image of the 1.3\,mm dust continuum emission, with contours
drawn at 10\,mJy beam$^{-1}$ ($10$\,$\sigma$) intervals, starting at 5\,mJy
beam$^{-1}$.  The synthesized beam is shown in the lower left corner.  (b) The
azimuthally averaged real ({\it black}) and imaginary ({\it gray}) components
of the 1.3\,mm continuum visibilities as a function of deprojected baseline
length.  (c) The broadband SED, with a composite photosphere model for the two
stars marked as a thin gray curve \citep[see][]{rosenfeld12}.  The thick gray
curve shows the {\it Spitzer} IRS spectrum.  (d) The velocity-integrated
intensities (0$^{\rm th}$ moment; contours) overlaid on the intensity-weighted
velocities (1$^{\rm st}$ moment; colors) for the $^{12}$CO $J$=2$-$1 line
emission.  Intensity contours start at 0.16\,Jy beam$^{-1}$ km s$^{-1}$ and
increase in 0.24\,Jy beam$^{-1}$ km s$^{-1}$ steps, and the color scale spans
an LSR velocity width of $\pm$3\,km s$^{-1}$ from the systemic value.  (e) The
same as (d), but for the $^{13}$CO $J$=2$-$1 line emission.  Intensity contours
are drawn at 0.15\,Jy beam$^{-1}$ km s$^{-1}$ intervals, starting at 0.10\,Jy
beam$^{-1}$ km s$^{-1}$.  (f) The integrated line profiles of the $^{12}$CO
({\it black}), $^{13}$CO ({\it red}), and C$^{18}$O $J=2-1$ ({\it blue}; not
detected) emission inside square regions 12\arcsec\ on a side.
\label{fig:summary}}
\end{figure}

A summarized representation of the SMA observations of V4046 Sgr is provided in 
Figure \ref{fig:summary}.  The synthesized map of 1.3\,mm dust continuum 
emission is shown in Figure \ref{fig:summary}(a), with contours starting at 
5\,mJy beam$^{-1}$ (5\,$\sigma$) and increasing at 10\,mJy beam$^{-1}$ 
(10\,$\sigma$) intervals.  The integrated flux density recovered in this map is 
$283\pm28\,$mJy (dominated by a 10\% systematic calibration uncertainty), 
consistent with a reasonable extrapolation from single-dish submillimeter 
photometry measurements \citep{jensen96}.  We find the mm-wave emission is 
concentrated in a bright, narrow ring centered at the mean stellar position 
\citep[$\alpha = 18^{\rm h}16^{\rm m}10\fs49$, $-32\degr47\arcmin34\farcs50$, 
J2000;][]{zacharias10}, with a double-peaked morphology due to limb brightening 
at the projected ring ansae \citep[see][]{andrews11}.  The apparent rotation 
between the semi-major axis of the disk and the line joining the two ansae 
is mostly an artifact of the $uv$ sampling and is reproduced in our models (see 
\S \ref{sec:model3}).  These emission peaks are separated by 0\farcs75 
($\sim$55\,AU) and have peak intensities of $50\pm5$\,mJy beam$^{-1}$ (dominated
by a 10\%\ calibration uncertainty; S/N $\approx$ 50).  The western peak appears
slightly brighter ($\sim$5\,mJy beam$^{-1}$) than its eastern counterpart, but a
proper evaluation of the significance of this discrepancy requires a more 
detailed analysis (see \S 5).  The emission ring itself is at best only 
marginally resolved, implying a width smaller than the 0\farcs4 ($\sim$30\,AU) 
minor axis of the synthesized beam.  Figure \ref{fig:summary}(b) shows the 
azimuthally-averaged profile of the real and imaginary continuum visibilities as
a function of deprojected baseline length, constructed assuming the disk viewing
geometry derived by \citet{rosenfeld12} ($i=33\fdg5$, PA$=76$\degr).  The 
visibility amplitudes exhibit the distinctive oscillation pattern that is 
characteristic of an emission ring, with nulls at $\sim$150 and 
350\,k$\lambda$.  The zero-spacing amplitude is estimated to be $\sim$315\,mJy, 
suggesting that roughly 10\%\ of the total flux density from the disk was 
filtered out of the synthesized map in Figure \ref{fig:summary}(a): presumably 
that emission is distributed on larger scales with low surface brightness.  

These SMA continuum measurements demonstrate that the V4046 Sgr binary joins 
the growing ranks of stars that host massive protoplanetary ``transition" disks 
\citep[e.g.,][]{andrews11}, with a central cavity of radius $\sim$29\,AU (see 
\S \ref{sec:model3}) that is substantially depleted of mm-sized dust particles. 
This classification is commensurate with the SED for V4046 Sgr, shown in Figure 
\ref{fig:summary}(c) and constructed from photometry in the literature 
\citep{hutchinson90,weaver92,jensen96,jensen97,rodriguez10,oberg11}, the 2MASS 
\citep{skrutskie06} and {\it IRAS} \citep{beichmann88} point source catalogs, 
and an archival {\it Spitzer} IRS spectrum.  The SED features the standard 
signature of a transition disk, with a distinctive ``dip" in the continuum near 
10\,$\mu$m suggesting that small, warm dust is preferentially depleted (although
{\it not} absent; see \S\ref{sec:model1}) near the central binary.  An accurate 
determination of the size of this putative cavity from the SED alone is not 
trivial \citep[e.g.,][]{calvet02}, although the very weak excess in the 
near-infrared makes it clear that the dust optical depths must be significantly 
diminished within at least a few AU of the central stars.  Note that the 
observed dip in the continuum covers a much wider wavelength range than was 
inferred by \citet{jensen97}.  Simulations of interactions between a stellar 
binary and its circumbinary disk suggest that the disk material should be 
inwardly truncated at a radius $\sim$2-5$\times$ the binary separation 
\citep[e.g.,][]{artymowicz94}: for V4046 Sgr, the disk truncation should occur 
at $\sim$0.2\,AU, much smaller than the continuum data suggest.  Therefore, the 
observed dust cavity is {\it not} related to interactions with the central 
binary.

\begin{figure}[t!]
\epsscale{0.95}
\plottwo{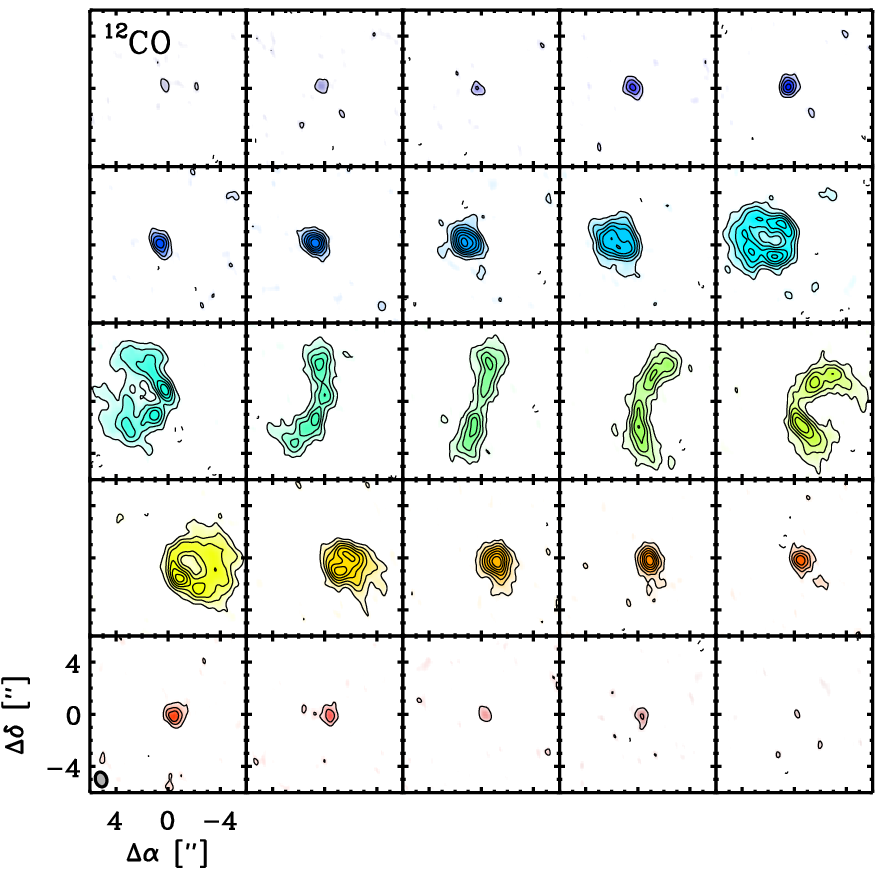}{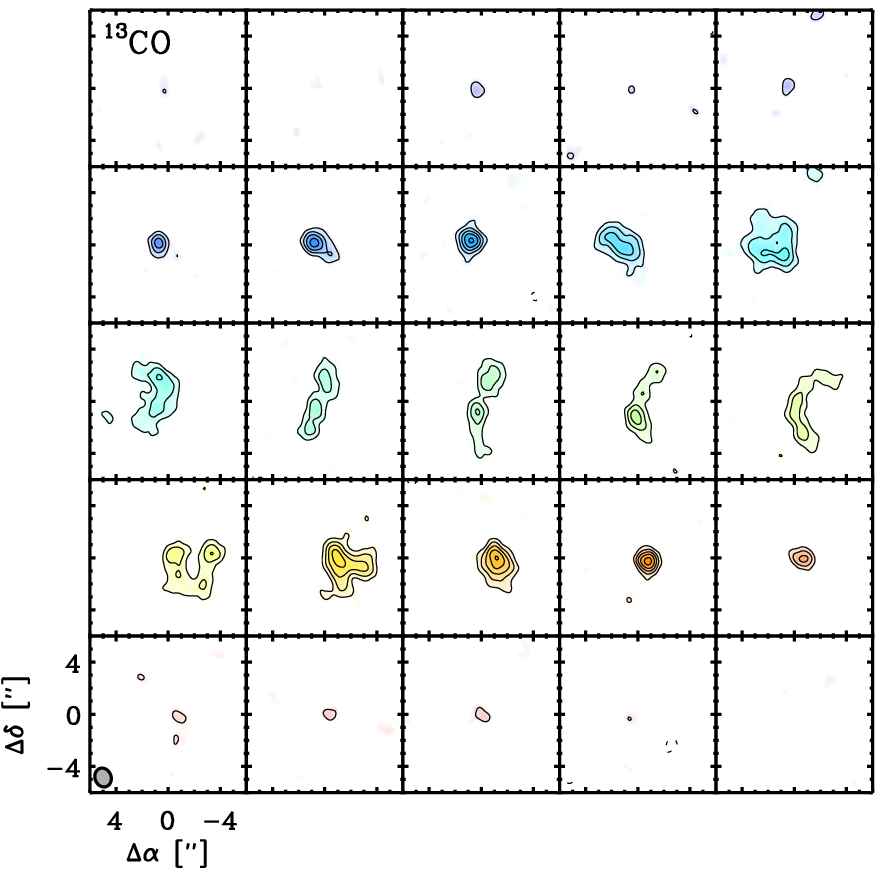}
\figcaption{Naturally weighted channel maps of the $^{12}$CO $J=2-1$ 
({\it left}) and $^{13}$CO $J=2-1$ ({\it right}) emission towards the V4046 
Sgr disk.  Channels are 0.4\,km s$^{-1}$ wide with the synthesized beam 
marked in the bottom left corner.  Contour levels are drawn at intervals of 
0.12 Jy\,beam$^{-1}$ ($3\,\sigma$).
\label{fig:chmaps}}
\end{figure}

The gas phase of the V4046 Sgr disk traced by CO isotopologue emission lines is
represented in Figures \ref{fig:summary}(d)-(f).  The $^{12}$CO and $^{13}$CO 
$J$=2$-$1 emission are displayed as moment maps in Figures \ref{fig:summary}(d) 
and (e).  The velocity-integrated intensities (0${\rm th}$ moment) are shown as 
contours at 3\,$\sigma$ intervals, starting at 2\,$\sigma$ (RMS noise levels 
are 0.08\,Jy beam$^{-1}$ km s$^{-1}$ for $^{12}$CO, 0.05\,Jy beam$^{-1}$ km 
s$^{-1}$ for $^{13}$CO), overlaid on the intensity-weighted velocities (1$^{\rm 
st}$ moment), with colors marking the velocity shift relative to line center.  
These maps exhibit the standard pattern of rotation, more clearly manifested in 
the individual channel maps shown together in Figure \ref{fig:chmaps}.  Line 
emission is firmly detected ($>$3\,$\sigma$, or 0.12\,Jy beam$^{-1}$ in each 
0.4\,km s$^{-1}$ channel) out to $\pm4.4$ or $\pm4.0$\,km s$^{-1}$ from the 
line center for $^{12}$CO and $^{13}$CO, respectively.  Given the V4046 Sgr 
stellar mass and disk inclination angle \citep[$i = 33\fdg5$, PA$ = 76$\degr 
with an ambiguity in the absolute orientation;][]{rosenfeld12}, those maximal 
projected velocities correspond to disk radii of $\sim$25-30\,AU, which is 
similar to the size of the dust cavity.  Therefore, these data do not have 
sufficient sensitivity in the line wings to rule whether or not there is 
CO inside the dust cavity.  The integrated line intensities derived from the 
0$^{\rm th}$ moment maps are $34.5\pm3.5$\,Jy km s$^{-1}$ for $^{12}$CO and 
$9.4\pm0.9$\,Jy km s$^{-1}$ for $^{13}$CO.  The peak intensities in the channel 
maps are $0.95\pm0.10$\,Jy beam$^{-1}$ ($22\pm 2$\,K; peak S/N $\sim$20) and 
$0.47\pm0.08$\,Jy beam$^{-1}$ ($6.5\pm0.8$\,K; peak S/N $\sim$11), respectively.
The C$^{18}$O emission is faint, and at best only marginally detected.  We 
estimate an integrated intensity of $\sim$0.6\,Jy km s$^{-1}$ from a 
0$^{\rm th}$ moment map, but suggest that this number be treated with caution: 
there is no firm detection in individual channel maps.  For reference, spatially
integrated spectra for the CO isotopologue lines are shown together in Figure 
1(f).

One of the notable features of Figure 1 is the disparity in the apparent 
spatial extents of the CO line and dust continuum emission.  The optically thick
$^{12}$CO line emission spans a diameter of 10\arcsec\ on the sky, corresponding
to a projected disk radius of $\sim$365\,AU.  Comparing Figures 1(a) and (d), we
find the dust ring extent is roughly 5$\times$ smaller.  The fainter, optically 
thinner $^{13}$CO emission still subtends $\sim$8\arcsec\ on the sky, 
$\sim$4$\times$ larger than the 1.3\,mm continuum.

\section{Modeling Analysis}

Having highlighted the basic observational characteristics of the gas and dust
tracers in the V4046 Sgr circumbinary disk, we now move to a more quantitative
effort to explore what the data can tell us about the disk structure.  We
concentrate our analysis on extracting a model for the radial distributions of
gas and dust based on the observational results described in \S 3.  To do that,
we first describe a flexible modeling framework for constructing disk 
structures given an arbitrary surface density profile and dust-to-gas ratio (\S
4.1), generalized from a similar analysis of the TW Hya disk 
\citep{andrews12}.  Using this formalism, we then consider in turn the SED 
and the resolved 1.3\,mm continuum observations and present model disk 
structures that optimally account for these data (\S 4.2.1 and 4.2.2) under the 
assumption of a constant dust-to-gas ratio.  After weighing the successes and 
failures of these two models, we relax the latter assumption and construct a 
model (\S 4.2.3) that is more consistent with all of the data: the broadband 
SED, the resolved dust continuum emission, and the resolved line emission from 
two CO isotopologues.  The ultimate success of this latter model provides some
insights on the shortcomings of our typical assumptions when analyzing disk
structures, which are discussed further in \S 5.

\subsection{Physical and Practical Overview}

Here we describe a generic prescription for constructing a parametric model of 
the two-dimensional density and temperature structure of both dust and gas in a 
protoplanetary disk, which builds on the previous work by 
\citet{andrews09,andrews10,andrews11,andrews12}.  We assume an axisymmetric,
two-dimensional model in a cylindrical reference frame with coordinates ($r$,
$z$).

Since dust grains dominate the opacity in a disk, we construct a model starting
with the distribution of its constituent solids.  In a general case with
different dust populations, the density distribution is simply defined as the
discrete sum $\rho_{\rm dust}(r,z) = \sum_j \rho_j(r,z)$, where
\begin{equation}
\rho_j(r,z) = \frac{\Sigma_j(r)}{\sqrt{2\pi} H_j(r)} \exp \left[-\frac{1}{2} \left(\frac{z}{H_j(r)}\right)^2\right]. 
\end{equation}
In Eq.~(1), $\Sigma_j(r)$ is the surface density profile and $H_j(r)$ is the
scale height profile for each dust population (labeled by index $j$).  To keep
the problem tractable, we consider only two dust populations in this study: a 
``midplane" population that dominates the disk mass and is composed of larger
grains that are vertically settled, and a less-abundant ``atmosphere" population
of smaller grains that is distributed to larger heights from the midplane.  We 
define parametric scale height profiles 
\begin{eqnarray}
H_{\rm atm}(r) &=& H_0 \,\, (r/r_0)^{\psi} \\
H_{\rm mid}(r) &=& \chi \,\, H_{\rm atm}(r),
\end{eqnarray}
where $H_{0}$ is the scale height at $r = r_0$, $\psi$ is the flaring index,
and $\chi$ is a scaling factor (in the range 0 to 1) that mimics dust 
settling: we assign a fixed $\chi = 1/2$ for simplicity.  The absorption and 
scattering opacities for each dust population were computed with a Mie code, 
assuming segregated spherical particles with the \citet{pollack94} mineralogical
composition and optical constants and a size distribution $n(a) \propto 
a^{-3.5}$ between 5\,nm and a maximum size (here $a$ is the grain radius).  In
this specific case, we fix $a_{\rm max}({\rm atm}) = 10$\,$\mu$m and
$a_{\rm max}({\rm mid}) = 1$\,cm, which reproduces the shape of the 
(sub)mm-wavelength SED well (see models described in \S\ref{sec:model1}, 
\ref{sec:model2}, and \ref{sec:model3}).

For any functional form for the surface densities, $\Sigma_j(r)$, the
two-dimensional distribution of dust densities (and optical depths, accounting
for the grain properties) can be specified with the above prescription.  Given
$\rho_{\rm dust}(r,z)$, we calculate the corresponding temperature structure,
$T_{\rm dust}(r,z)$, using the Monte Carlo radiative transfer code {\tt
RADMC-3D}\footnote{\url{http://www.ita.uni-heidelberg.de/$\sim$dullemond/software/radmc-3d/}}, assuming an incident radiation field from the central binary.
For the latter, we use a composite spectrum of two \citet{lejeune97} model
photospheres with \{$T_{\rm eff,1} = 4350$\,K, $L_{\ast,1} = 
0.35$\,L$_{\odot}$, $M_{\ast,1} = 0.90$\,M$_{\odot}$\} and \{$T_{\rm eff,2} = 
4060$\,K, $L_{\ast,2} = 0.25$\,L$_{\odot}$, $M_{\ast,2} = 0.85$\,M$_{\odot}$\} 
\citep[for details, see][]{rosenfeld12}.

To construct a corresponding gas disk structure, we define a vertically 
integrated dust-to-gas mass ratio $\zeta$ that may vary spatially, such that 
$\Sigma_{\rm gas}(r) = \Sigma_{\rm dust}(r)/\zeta(r)$, where $\Sigma_{\rm 
dust}(r) = \sum_j \Sigma_j(r)$.  The two-dimensional density structure of the 
gas, $\rho_{\rm gas}(r, z)$, is determined by numerically integrating the 
differential equation that describes vertical hydrostatic equilibrium for a 
given gas temperature distribution, $T_{\rm gas}(r, z)$ \citep[see][their 
Eq.~5]{andrews12}.  At the midplane ($z = 0$), the gas and dust are assumed to 
be thermally coupled: $T_{\rm gas}(r, z = 0) = T_{m} = T_{\rm dust}(r, z = 
0)$.  At larger vertical heights, the gas temperatures are permitted to deviate 
from the dust temperatures.  To facilitate those departures, we adopt the 
parameterization introduced by \citet{dartois03},
\begin{equation}
 T_{\rm gas}(r, z) = \left\{
\begin{array}{ll}
T_a + \left(T_m - T_a\right) \left[\cos{\frac{\pi z}{2 z_q}}\right]^{2\delta} & \mbox{if $z < z_q$} \\
T_a & \mbox{if $z \ge z_q$} 
\end{array},
\right. 
\end{equation}
where $T_a = T_{a,0} (r/r_0)^{-q}$ is a parametric temperature profile for the
gas in the disk atmosphere, which is explicitly defined here as heights larger
than a fiducial value, $z_q$.  We fix $z_q = 2 H_{\rm gas} = 2 c_s/\Omega$,
where $H_{\rm gas}$ is the hydrostatic gas scale height evaluated at the
midplane temperature ($T_m$), defined as the ratio of the sound speed ($c_s$)
to the Keplerian angular velocity ($\Omega$).  The index parameter $\delta$
determines the shape of the vertical temperature gradient of the gas, and is 
fixed here to $\delta = 2$, following \citet{dartois03}.  Finally, to insure a 
physically realistic model, we impose the additional criterion that the gas 
cannot be colder than the dust: if $T_{\rm gas}(r, z)$ calculated using Eq.~(4) 
is less than $T_{\rm dust}(r, z)$ computed from the radiative transfer 
simulation, we set $T_{\rm gas} = T_{\rm dust}$.  In practice, this formulation
provides a straightforward means of permitting extra gas heating relatively 
near the star, while (if desired) maintaining thermal coupling between the gas
and dust at all heights at large disk radii ($r \gtrsim 100$\,AU; see \S 4.2).
For the kinematic structure of the disk, we assume the gas is in Keplerian
rotation around the central stars, 
${v_{\theta}}^2=GM_\ast/r$, 
and that the emission line profile widths are determined from the quadrature sum
of a thermal and turbulent broadening term, with a fixed and constant value for 
the latter of $\xi = 0.1$\,km s$^{-1}$.

Having defined the gas disk structure, we need to assign an abundance
distribution for the tracer molecule, CO.  Following \citet{aikawa99}, we
assume that 80\%\ of the gas is composed of H$_2$ and then assume a {\it
constant} fractional abundance $f_{\rm co} = n(^{12}$CO)/$n({\rm H}_2)$ such
that
\begin{equation}
n(^{12}{\rm CO}) = 0.8 f_{\rm co} \frac{\rho_{\rm gas}(r, z)}{\mu m_{\rm H}}
\end{equation}
in the region of the disk we term the ``abundant layer".  In Eq.~(5), 
$\mu$ is the mean molecular weight of the gas ($\mu = 2.37$) and $m_{\rm H}$ is 
the mass of a hydrogen atom.  The abundant layer is defined based on the 
treatment of \citet{qi08,qi11}, which assumes CO is present in the gas phase if:
(1) the local gas temperatures are high enough that the CO is not frozen onto 
the dust grains, and (2) the vertically-integrated column density is higher than
the penetration depth of photodissociating radiation.  At each radius,
Eq.~(5) is used to compute the CO densities if
\begin{eqnarray}
T_{\rm frz} &\le& T_{\rm gas}(r, z) \\
\sigma_s &\le& f_{\rm H} \int_z^{\infty} n_{\rm gas}(r, z^{\prime}) \,\, dz^{\prime} 
\end{eqnarray}
are valid: here, $T_{\rm frz}$ is the CO freezeout temperature, $\sigma_s$ is
the photodissociation column, and $f_{\rm H}$ is the fraction of H nuclei in 
the gas.  Guided by \citet{aikawa99} and \citet{qi11}, we assume $f_{\rm H} = 
0.706$ and fix $T_{\rm frz} = 19$\,K and $\sigma_s = 
5\times10^{20}$\,cm$^{-2}$.  At locations where Eqs.~(6) and (7) are not met, 
we sharply (and arbitrarily) reduce the $^{12}$CO abundance by scaling 
$f_{\rm co}$ down by a factor of $10^8$.  The CO isotopologue abundances are 
assumed to follow $^{12}$CO with the interstellar medium ratios inferred by 
\citet{wilson99}: $n(^{12}$CO)/$n(^{13}$CO$) = 69$ and 
$n(^{12}$CO)/$n($C$^{18}$O$) = 557$.

The formalism described above and encapsulated in Eqs.~(1)-(7) fully defines the
dust and gas structure of a model disk.  In practice, the dust structure 
depends on some description of the surface density profiles for each dust 
species, $\Sigma_j(r)$, and their corresponding vertical height profiles, 
$H_j(r)$: here, the latter is fully specified with two parameters, \{$H_0$, 
$\psi$\}.  Likewise, the gas structure is determined by a dust-to-gas mass ratio
profile $\zeta(r)$, and three additional free parameters: \{$T_{a,0}$, $q$\} to 
define the atmosphere temperature profile and \{$f_{\rm co}$\} to assign the CO 
molecular abundance.  For a given model, we generate synthetic data products to 
compare with the observations and evaluate our parameter choices and the 
underlying model characterization.  For the dust, we use the ray-tracing 
capability of {\tt RADMC-3D} to produce a synthetic SED and set of 1.3\,mm
continuum visibilities sampled at the same spatial frequencies as observed by
the SMA.  For the gas, we use the molecular excitation and line radiation 
transfer code {\tt LIME} \citep{brinch10} to calculate the non-local 
thermodynamic equilibrium level populations based on information from the LAMDA
database \citep{schoier05} and then calculate synthetic spectral visibilities
for each of the CO isotopologues in the velocity channels and spatial 
frequencies sampled by the SMA.  In all models, we assume the disk inclination
($i = 33\fdg5$), major axis position angle (PA = 76\degr), total stellar mass
($M_{\ast} = 1.75$\,M$_{\odot}$), and disk center position derived by
\citet{rosenfeld12}, as well as the distance ($d = 73$\,pc) inferred by
\citet{torres06,torres08}.

\subsection{Model Results}

The modeling framework we have constructed is relatively complex, with some
fairly severe parameter degeneracies \citep[for a more detailed discussion,
see][]{andrews09,andrews11,andrews12,qi11}.  Moreover, the computational
effort required for the radiative transfer and non-LTE excitation calculations
remains costly, severely limiting our capability to use either stochastic or
deterministic optimization algorithms to robustly probe the parameter-space.
In practice, we build structure models manually, using small explorations of
each parameter to reproduce the key observational features of interest.  This
means that the models presented below are not quantitatively optimized, nor
unique.  However, they are still of great interest since they serve as
qualitative illustrations of key features in the V4046 Sgr disk structure that
are not otherwise easily accessible.  In the following sections (\S 4.2.1 and
4.2.2), we aim to provide a pedagogical guide to the logic behind the more
complex disk structure we ultimately will advocate for the V4046 Sgr disk (\S
4.2.3).

\subsubsection{Model to Reproduce the SED}\label{sec:model1}

\begin{figure}[t!]
\epsscale{0.95}
\plotone{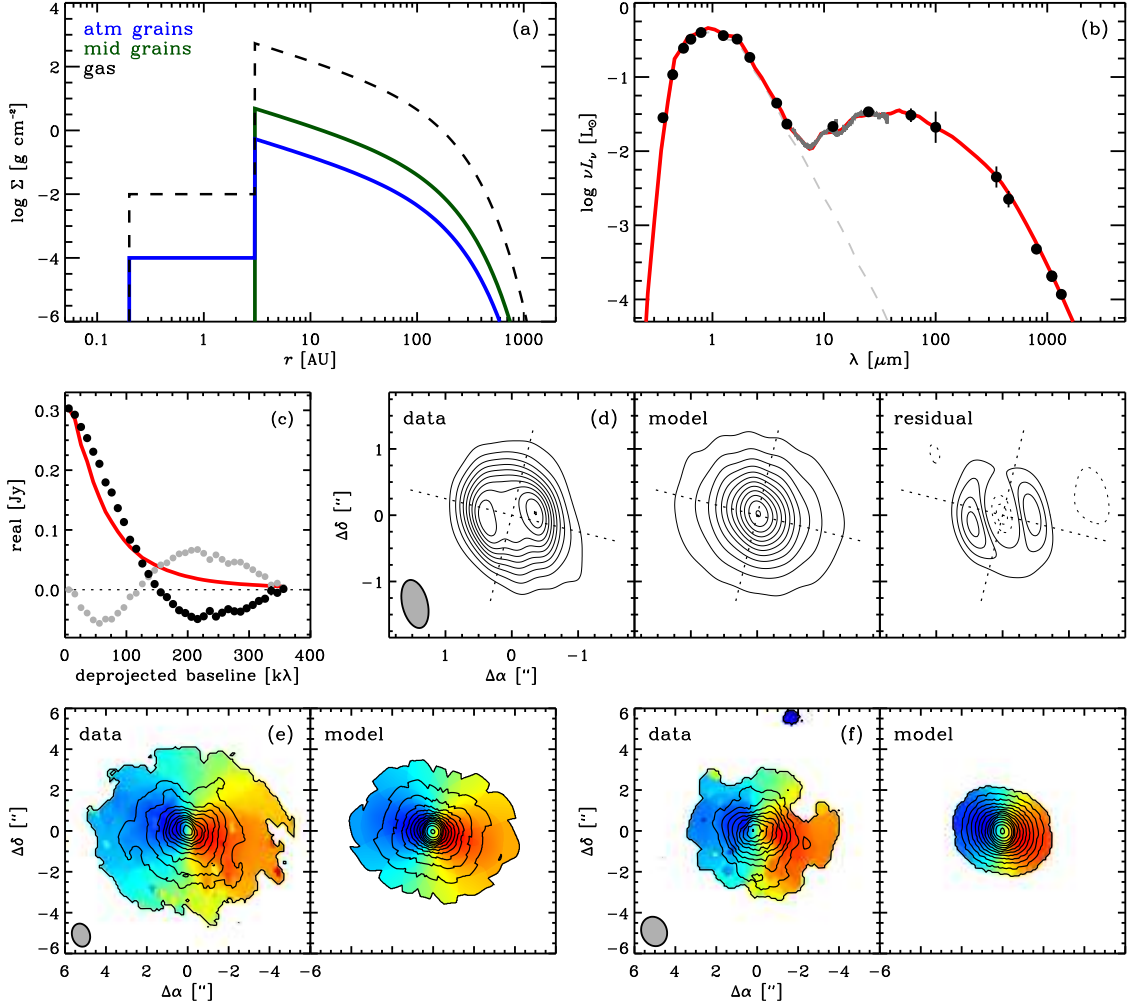}
\figcaption{Results for the modeling effort focused on the SED (\S 4.2.1).  (a)
Surface density profiles for the gas ({\it dashed}), the large ``midplane"
({\it green}) and small ``atmosphere" ({\it blue}) grain populations.  (b) The
SED ({\it black points} and {\it solid gray curve}) compared with the model
({\it red}).  The stellar photosphere model is shown as a dashed gray curve.
(c) The deprojected, azimuthally-averaged (real) 1.3\,mm continuum visibility
profile, with model overlaid (in {\it red}); gray points mark the residuals.
(d) The observed, simulated, and residual 1.3\,mm emission synthesized maps.
Contours are drawn at 5\,mJy beam$^{-1}$ (5\,$\sigma$) intervals.  Note the
smaller map size compared to Figure \ref{fig:summary}.  (e) A comparison of the
observed and simulated $^{12}$CO moment maps.  (f) The same as panel (e), but
for the $^{13}$CO line.  Contour levels and map sizes for panels (e) and (f)
are the same as in Figure \ref{fig:summary}.  \label{fig:modsum1}}
\end{figure}

As a starting point, we focus on building a structure model that is designed to
reproduce the broadband SED of V4046 Sgr, with some input from the CO
observations to guide our description of the gas disk structure.  Following the
modeling prescription for transition disks described by \citet{andrews11}, we
adopt a truncated version of the \citet{lynden-bell74} similarity solution for
a thin, Keplerian accretion disk with a time-independent, power-law viscosity
profile to describe the surface densities of the dust \citep[see
also][]{hartmann98},
\begin{equation}
 \Sigma_{\rm dust}(r) = \left\{
\begin{array}{ll}
\Sigma_{ss}(r) = \Sigma_c \left(\frac{r}{r_c}\right)^{-\gamma} \exp \left[-\left(\frac{r}{r_c}\right)^{2-\gamma}\right]  & \mbox{if $r \ge r_{\rm cav}$} \\
\Sigma_{\rm cav} & \mbox{if $r_{\rm in} \le r < r_{\rm cav}$} 
\end{array},
\right. 
\end{equation}
where $r_c$ is a characteristic radius, $\gamma$ is an index parameter, 
$\Sigma_c = e \cdot \Sigma_{\rm dust}(r_c)$, $r_{\rm cav}$ is a ``cavity" 
radius, $\Sigma_{\rm cav}$ is a constant, and $r_{\rm in}$ is the inner edge of
the model.  Both grain populations were forced to follow the behavior in 
Eq.~(8): outside the disk cavity, 90\%\ of the mass was apportioned to the
midplane grains ($\Sigma_{\rm mid} = 0.9\,\Sigma_{\rm dust}$, $\Sigma_{\rm atm} 
= 0.1\,\Sigma_{\rm dust}$, for $r \ge r_{\rm cav}$), but the tenuous material
inside the cavity was assumed to be all atmosphere grains ($\Sigma_{\rm mid} = 
0$, $\Sigma_{\rm atm} = \Sigma_{\rm cav}$, for $r < r_{\rm cav}$).  Since the
SED is composed of unresolved photometric measurements, it has no ability to
constrain the gradient $\gamma$.  Therefore, we fixed $\gamma = 1$, a typical
value for disks \citep{andrews09,andrews10}.  The inner radius of the model was
set to $r_{\rm in} = 0.2$\,AU, the outer edge of the zone expected to be
cleared by dynamical interactions with the central binary
\citep{artymowicz94}.  For the gas phase, we assumed a standard, spatially 
uniform dust-to-gas ratio, $\zeta(r) = \zeta = 0.01$ 
\citep{pollack94,dalessio01}.

After some iteration between dust and gas structures, we identified a model
that can reproduce well both the broadband SED and SMA observations of CO line
emission: the results are highlighted in Figure \ref{fig:modsum1}.  Using the 
morphology of the continuum ``dip" in the infrared SED, we inferred a cavity 
radius $r_{\rm cav} = 3$\,AU, roughly 15$\times$ larger than could be explained 
by tidal stripping from the V4046 Sgr binary.  As with most other transition 
disks, this cavity is not empty: the weak infrared excess (and weak silicate 
emission feature) noted in Fig.~\ref{fig:modsum1}b is accommodated with 
$\Sigma_{\rm cav} \approx 10^{-4}$\,g cm$^{-2}$.  In the mid-infrared, the 
excess spectrum was fit with a scale height of $H_0 = 0.4$\,AU at $r_0 = 
10$\,AU and a flaring index $\psi = 1.25$ \citep[although the latter is not 
well-determined in transition disks; see][]{andrews11}.  The total mass of the 
dust structure (the integral of Eq.~8 over the disk area) was $M_{\rm dust} 
\approx 9\times10^{-4}$\,M$_{\odot}$, determined from the luminosity of the 
millimeter-wave SED.  The characteristic size of the disk was estimated to be 
$r_c = 75$\,AU, based on the extent of the $^{12}$CO emission.  Those data were 
described well with a CO abundance $f_{\rm co} = 3\times10^{-6}$ and a steep 
atmospheric temperature profile with $T_0 = 200$\,K (at $r_0 = 10$\,AU) and 
$q = 0.8$.  In practice, this $T_a(r)$ profile means the gas and dust are 
thermally coupled for $r \gtrsim 150$\,AU, but there is some source of 
additional gas heating in the disk atmosphere at smaller radii.  One possible 
source for that heating is the strong X-ray emission from the central binary 
(see \S\ref{sec:discussion}).

Although by design this model provides a good match to the SED and CO emission,
it should be obvious from Figure \ref{fig:modsum1} that it is
irreconcilable with the resolved 1.3\,mm continuum data.  However, these 
failures are instructive in refining the model.  The residuals in Figures 
\ref{fig:modsum1}(c)-(d) point out two key issues.  First, the cavity size 
derived from the SED alone substantially under-estimates the size that can be 
{\it directly} measured from the resolved 1.3\,mm data.  In \S 3, we estimated 
a cavity radius of $\sim$29\,AU (and will quantify that further below), 
$\sim$10$\times$ larger than derived here.  This kind of discrepancy is not 
uncommon for transition disks \citep[e.g., DM Tau;][]{andrews11}: any inference 
of a size scale from unresolved observations is inherently uncertain and may 
be a signature of how the dynamics of grains depends upon their size (see 
\S\ref{sec:discussion}).  Second, 
the spatial extent of this model is much larger than the 1.3\,mm continuum 
distribution.  This latter point is manifested in the poor match to the 
visibility profile in Figure \ref{fig:modsum1}(c) on large spatial scales 
(short baselines), where the model is clearly more resolved than the data 
permit (the negative residuals outside the emission ring in Figure 
\ref{fig:modsum1}(d) represent the same aspect of this model failure).  Since 
$r_c$ was determined from the CO data, this discrepancy is not surprising: we 
already highlighted in \S 3 how the CO emission is much more extended than the 
dust emission (see Fig.~\ref{fig:summary}).

\subsubsection{Model to Reproduce the Resolved 1.3\,mm Continuum Emission}
\label{sec:model2}

\begin{figure}[t!]
\epsscale{0.95}
\plotone{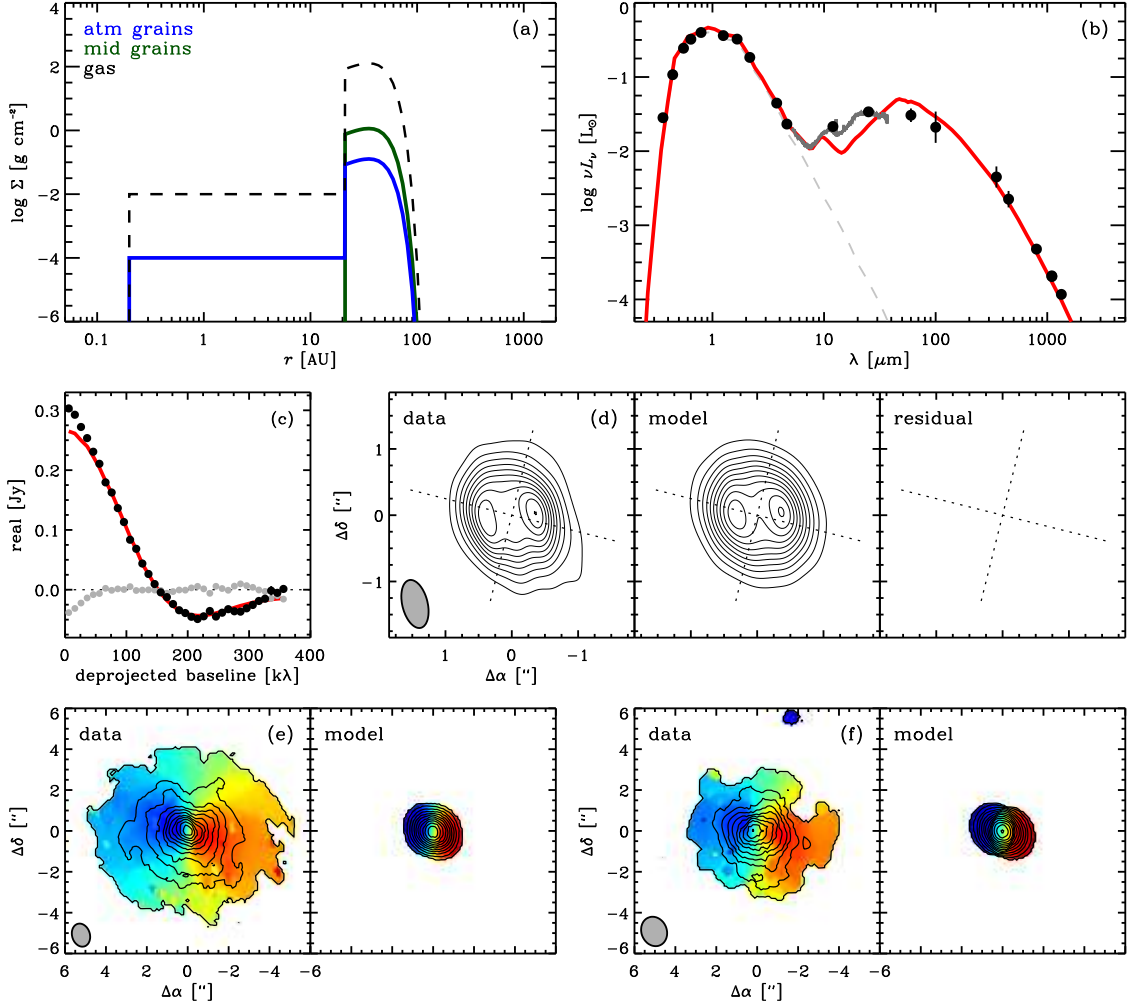}
\figcaption{Results for the modeling effort focused on the resolved 1.3\,mm 
dust continuum emission (first model described in \S 4.2.2).  See Figure 
\ref{fig:modsum1} for a description of the contents of individual panels.  
\label{fig:modsum2a}}
\end{figure}

\begin{figure}[t!]
\epsscale{0.95}
\plotone{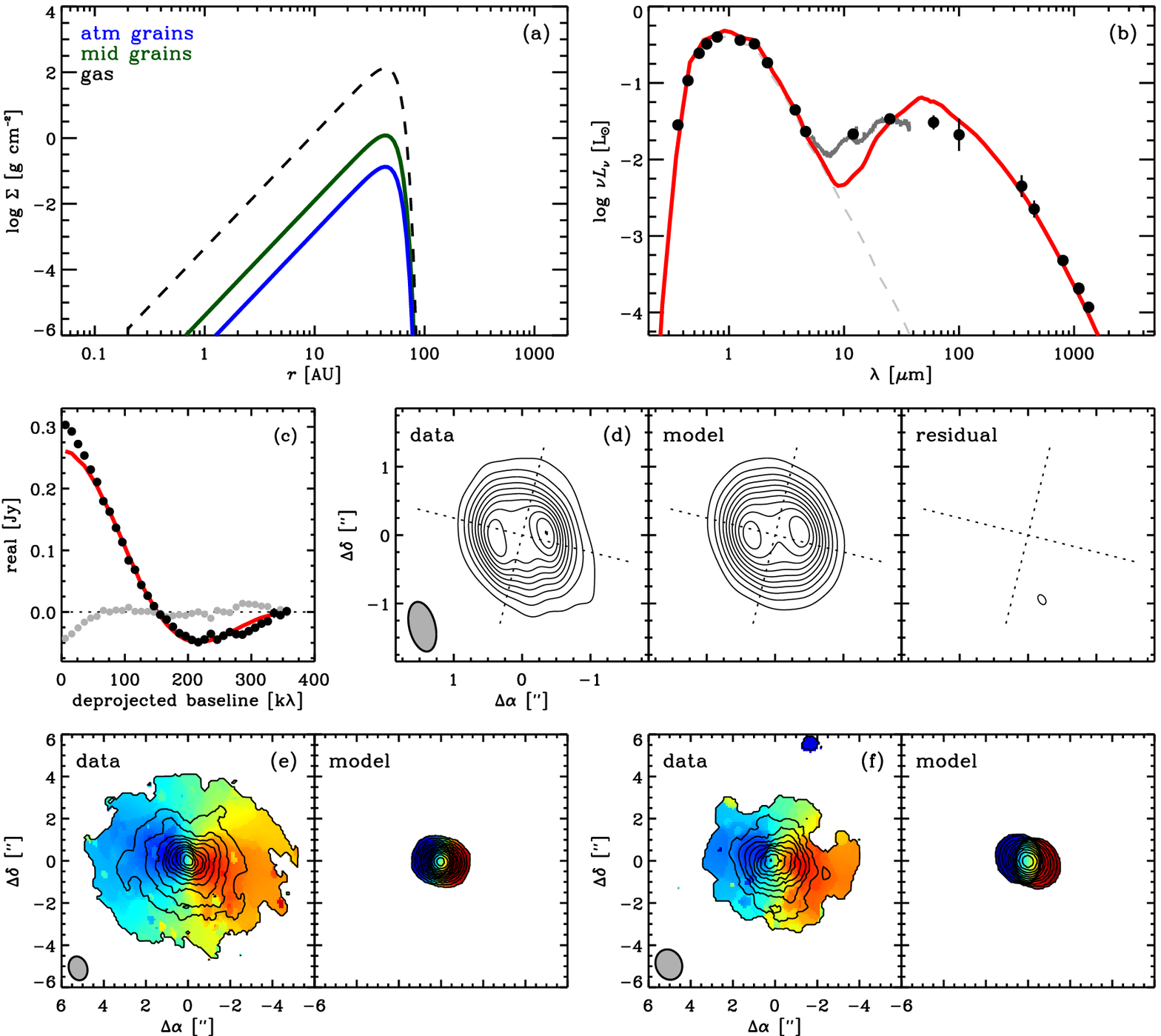}
\figcaption{Results for the modeling effort focused on the resolved 1.3\,mm 
dust continuum emission (second model described in \S 4.2.2).  See Figure 
\ref{fig:modsum1} for a description of the contents of individual panels.  
\label{fig:modsum2}}
\end{figure}

Informed by the failures of the previous model, we shift the focus here to
building a model that is more consistent with the resolved morphology of the 
1.3\,mm dust emission.  Two prescriptions for the surface density profile are 
presented, both of which can reproduce the size and shape of the mm-wave 
emission.  Considering the visibility profile (Fig.~\ref{fig:summary}a), the former 
is constrained by the short baselines 
($\mathcal{R}\lesssim150$\,k$\lambda$) and the latter is dictated by the 
position of the null and deep trough at higher spatial frequencies.  The results
shown in Figure \ref{fig:modsum2a} are for the same relative ratios of midplane 
and atmosphere grains and model parameterization defined in \S\ref{sec:model2}, 
but with the the cavity size, $r_{\rm cav}$, enlarged to 21\,AU (i.e., adjusted
without regard to the infrared SED).  In addition, 
this model has $\gamma=-1.5$, $r_c=45$\,AU, $\psi = 1.5$, 
$H_0 = 0.45$\,AU at $r_0 = 10$\,AU, and 0.0011\,$M_\odot$ of dust.  Note that a 
negative $\gamma$ implies a density profile that {\it rises} with 
radius out to $r_c$, but then is sharply truncated.  In essence, it naturally
produces a relatively narrow ``ring" of material \citep{isella09}, making it a
practical means of describing the resolved morphologies of transition disks 
\citep[e.g.,][]{isella10,isella12,andrews11b,brown12}.  We also 
consider an alternative parameterization of the surface density profile where 
the viscous disk similarity solution, $\Sigma_{\rm dust}(r) = \Sigma_{ss}(r)$ 
(see Eq.~8), is valid for all $r$. Figure \ref{fig:modsum2} summarizes this 
second model which features a large, negative $\gamma = -3.5$ with 
$r_c = 48$\,AU, $\psi = 1.5$, $H_0 = 0.4$\,AU at $r_0 = 10$\,AU and 
0.0010\,$M_\odot$ of dust.

Although both of these models reproduces well the morphology of the 1.3\,mm 
emission ring (Figs.~\ref{fig:modsum2a},\ref{fig:modsum2}c,d), they each 
fail to account for the observations of the V4046 Sgr disk in three similar, 
illuminating ways.  We also experimented with alternative prescriptions for the
surface densities (e.g., a power-law with an outer edge cut-off), and found 
qualitatively similar results.  First, they produce too
deep and wide of a ``dip" in the infrared SED in Figures 
\ref{fig:modsum2a},\ref{fig:modsum2}(b), because there is not enough dust 
interior to the density peak.  Note that the mid/far-infrared
excess of the models is a manifestation of this lack of inner disk material: in 
the absence of an inner disk, a larger surface area of more distant material in 
the model ring can be directly illuminated by the central stars, producing a 
``bump" in the continuum at $\sim$50\,$\mu$m.  Second, if we assume a standard, 
uniform dust-to-gas ratio as before ($\zeta = 0.01$), there is not enough gas 
outside the density peak in either of these models to account for the observed 
spatial extent of the CO emission 
(Figs.~\ref{fig:modsum2a},\ref{fig:modsum2}e,f).  For a constant $\zeta$, this 
size discrepancy is present regardless of the assumed CO abundance or gas 
temperature distribution, although the results in Figures \ref{fig:modsum2a} and
\ref{fig:modsum2} use the same $f_{\rm co}$ and $T_a(r)$ that were adopted in 
\S 4.2.1.  And third, the models do not quite generate enough 1.3\,mm dust 
emission on large angular scales (deprojected baselines $\le$50\,k$\lambda$; 
see Figs.~\ref{fig:modsum2a},\ref{fig:modsum2}c).  The origins of 
this last discrepancy are indeed subtle.  We are unable to account for all the 
emission through a simple scaling of the dust mass because the model ring is 
already nearly optically thick (this is exacerbated by our assumption of 
isotropic scattering off large, spherical grains).  Increasing the densities 
further effectively impedes the
radiation transfer to the midplane, making the dust cooler and actually {\it 
decreasing} the 1.3\,mm luminosity.  Instead, we interpret this emission 
mismatch as an intrinsic model deficiency: the addition of a low-density, 
spatially-extended dust component could reconcile the model visibilities with 
the data.  Note that this faint ($\sim$10-15\%\ of 
the total flux) emission ``halo" would be strongly spatially filtered in the 
synthesized images, as was found for the SMA observations in \S 3.

\subsubsection{A Hybrid Model to Reproduce All the Data}\label{sec:model3}

The three different models constructed above illustrate the key features of the 
V4046 Sgr disk structure.  To review those results, we have identified three 
basic structural elements that are required in any model prescription that aims 
to successfully reproduce the observations:
\begin{enumerate}
\item The vast majority of the large grain population must be strongly
concentrated in a {\it narrow} ring with a large central cavity to explain the
1.3\,mm continuum visibilities;
\item Inside that cavity, a reservoir of small dust particles (which produce
little mm-wave emission) at radii of a few AU are required to account for the
morphology of the infrared SED;
\item A large-scale, extended halo structure is necessary to explain the CO
line emission size and a faint, strongly spatially filtered 1.3\,mm continuum 
component, but there appears
to be substantially less mass in the large dust particles relative to the gas
outside the concentrated ring component.
\end{enumerate}
Here, our goal is to combine the successful aspects of the models described
above into a hybrid structure that incorporates these three elements and is
commensurate with all of the observations.

We define a surface density profile for the dust that is composed of three
individual components, each corresponding to one of the structural elements
enumerated above:
\begin{equation}
\Sigma_{\rm dust}(r) = \Sigma_{\rm ring}(r) + \Sigma_{\rm in}(r) + \Sigma_{\rm halo}(r).
\end{equation}
We elect to use simple Gaussian profiles for the distribution of material in the
ring and inner disk components,
\begin{equation}
\Sigma(r) = \frac{\Sigma_0}{\sqrt{2\pi}\sigma} \exp \left[-\frac{1}{2}\left(\frac{r-\mu}{\sigma}\right)^2\right],
\end{equation}
where we define parametric constants that correspond to peak radii \{$\mu_{\rm 
ring}$, $\mu_{\rm in}$\}, profile widths \{$\sigma_{\rm ring}$, $\sigma_{\rm 
in}$\}, and normalizations \{$\Sigma_{0, {\rm ring}}$, $\Sigma_{0, {\rm in}}$\}
for both structure components separately.  We assume that the ring component is
composed solely of the large ``midplane" grains, and that the inner component
is made up entirely of the small ``atmosphere" grains.  For the halo component,
we again assume a truncated similarity solution model (as in \S 4.2.1), but
eschew the sharp inner edge for one with a smoother, Gaussian taper:
\begin{equation}
\Sigma_{\rm halo}(r) = \Sigma_{ss}(r; \{\Sigma_c, r_c, \gamma\}) \, \exp \left[-\left(\frac{\mu_{\rm ring}}{r}\right)^2\right].
\end{equation}
The functional form of Eq.~(11) is arbitrary, and was selected entirely for
purposes of convenience in the radiative transfer modeling (tying the
exponential turnover scale to the ring component center makes a joint
exploration of the model space less cumbersome).  We fix $\gamma = 1$ as before 
and use a 6:1 mass ratio between the midplane and atmosphere grains.  The gas is
assumed to follow the small grains in both the halo and inner disk components
separately -- i.e. decoupled from the large grain population -- each with a 
ratio $\zeta$.  We use the halo gas profile as an upper limit on the surface 
density of midplane grains in the ring. For radii where the ratio between the 
two would exceed unity (e.g. in the space between the two Gaussian 
distributions), we reduce $\Sigma_{\rm ring}$ to equal the gas surface density 
from the halo component.  Either increasing the gas or reducing the dust density
has no significant effect upon our observables since the subtended region is 
very narrow and the densities are already low.

\begin{figure}[t!]
\epsscale{0.95}
\plotone{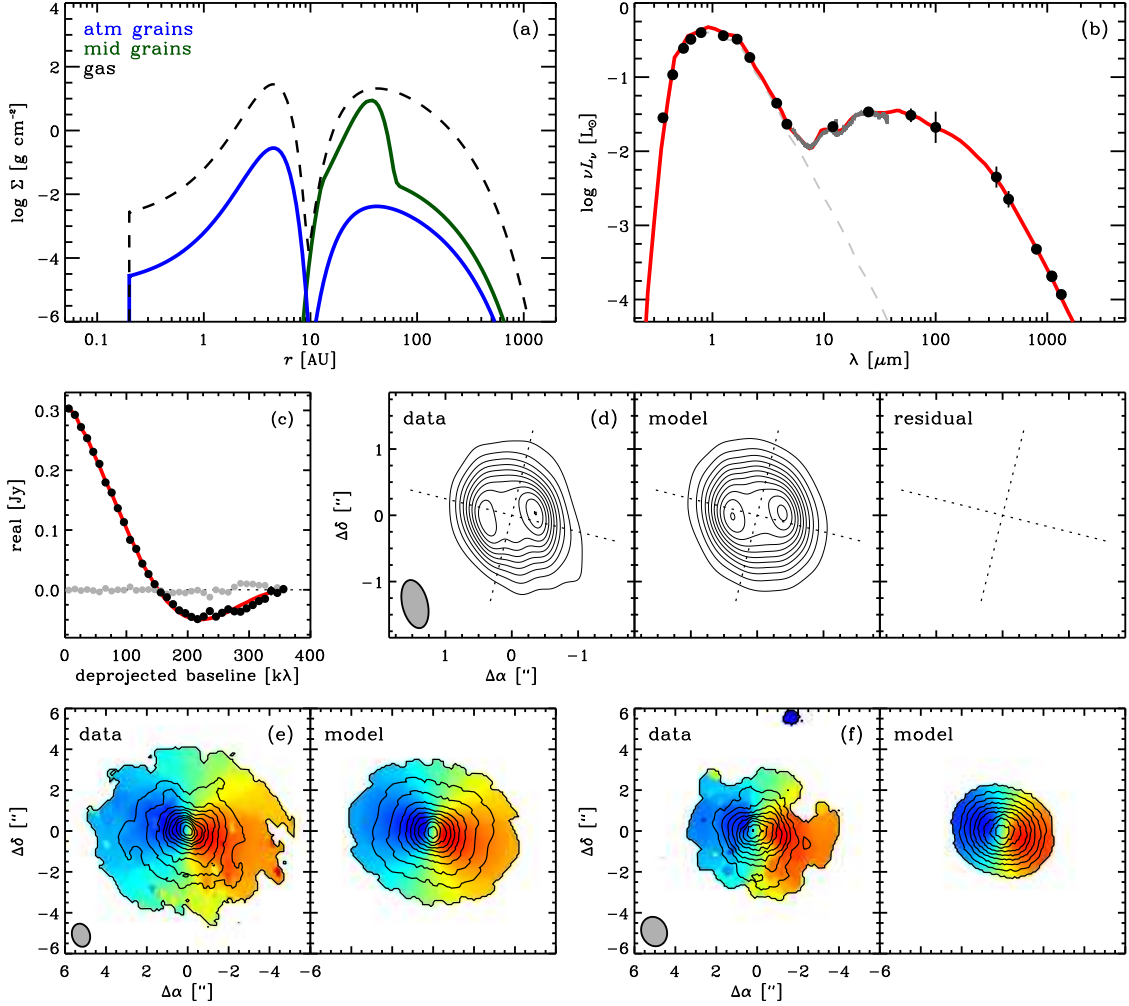}
\figcaption{Results for the modeling effort that successfully reproduces all of 
the data (SED, 1.3\,mm continuum visibilities, and CO spectral images).  See 
Figure \ref{fig:modsum1} for descriptions of individual panels.  
\label{fig:modsum3}}
\end{figure}

A model constructed with the recipe outlined above is superior to those 
presented in \S\ref{sec:model1} and \ref{sec:model2} in terms of reproducing all
the relevant observations of the V4046 Sgr transition disk, as
demonstrated in Figure \ref{fig:modsum3}.  In this case, we assumed that the 
same dust scale height distributions used in the previous sections were 
applicable to all of the structure elements.  For the ring component, we 
identify its center at $\mu_{\rm ring} = 37$\,AU and width $\sigma_{\rm ring} = 
7$\,AU (corresponding to a FWHM of 16\,AU, or 0\farcs2 projected on the sky; 
notably smaller than the SMA angular resolution), and allocate a large total 
dust mass of 0.004\,M$_{\odot}$.  The representative size of the 
central cavity is estimated to be $29$\,AU by subtracting one half of the FWHM 
from the ring radius.  In the inner disk, we find 
$\mu_{\rm in} = 4$\,AU, $\sigma_{\rm in} = 1$\,AU, and a total dust mass of
$2\times10^{-5}$\,M$_{\odot}$.  To account for the faint, extended 1.3\,mm 
emission component and the size of the CO emission, a halo component with a 
characteristic radius $r_c = 75$\,AU and total dust mass of 
$\sim$10$^{-4}$\,M$_{\odot}$ is sufficient.  To produce sufficient CO line 
intensities for the same gas temperature and molecular abundance structures
highlighted in \S 4.2.1, we set $\zeta = 0.0014$ in the outer disk, 
{\it relative to the halo component only}.  The weak C$^{18}$O emission of this 
model is consistent with the observations (i.e. a non-detection).  In the inner 
disk, we have kept 
$\zeta = 0.01$ fixed as before.  The total (gas+dust) mass in this model is 
$\sim$0.094\,M$_{\odot}$ and the disk-integrated dust-to-gas mass ratio is 
$\sim$0.047.

\clearpage

\section{Discussion}\label{sec:discussion}

We have presented sensitive, high angular resolution SMA observations of the 
1.3\,mm dust continuum and spectral lines of three CO isotopes from the V4046 
Sgr circumbinary disk. We have modeled the dust and gas structures and 
identified three fundamental signatures of profound evolution in this disk.  
First, the majority of the large grains are distributed in a narrow ring 
centered creating a large central cavity ($r\sim29$\,AU).  Second, there exists 
a significant population of $\mu$m-sized grains that have not been cleared from 
the center of this cavity.  Third, the gas disk extends much farther than the 
compact dust ring, suggesting that this disk has a radial variation in the 
dust-to-gas mass ratio.

To explain the striking discrepancy between the size of the gas and dust 
distributions, we invoke a model with a strong radial variation in the relative 
mass ratio of the large, midplane grains and the gas.  The dust-to-gas mass
ratio in the model is large at the dust ring ($\zeta \sim 0.1$ -- $1$) and small
outside of it ($\zeta \sim 10^{-3}$).  Similar discrepancies between the gas and
dust distributions have been noted for other disks, such as those around IM Lup 
\citep{panic08}, TW Hya \citep{andrews12}, and LkCa 15 \citep{isella12}.
While our preferred model does feature an extended, but tenuous, dusty halo, 
models where the dust uniformly follows the gas disk either produce too much 
continuum or too little CO emission at large radii and clearly disagree with 
the observations (see \S\ref{sec:model1} and \S\ref{sec:model2}).  We caution 
that the model $^{12}$CO and $^{13}$CO line emission is optically thick and so 
the derived dust-to-gas mass ratios are upper limits.  Additionally, the dust 
mass in the model is sensitive to the dust opacities and so other values of the
dust-to-gas mass ratio and total disk mass are possible. The precise form and 
normalization of $\zeta(r)$ are model dependent, but the observations seem to 
suggest that it is a strongly decreasing function.

While the underlying physical cause of the discrepancy between the size of the 
gas and dust distributions is not clear, there is a 
natural explanation in the growth and inward radial drift of dust particles 
\citep{weidenschilling80,brauer08a,birnstiel10}.  Drift is a fundamental feature
in the migration of disk solids, generated because the radial pressure gradient
in the disk causes the gas to rotate at slightly sub-Keplerian velocities while 
the dust particles are on Keplerian orbits.  The resulting headwind creates a 
drag-force on the dust that relates to the size of the particle, preferentially
moving the large particles towards the central star \citep{weidenschilling77,
takeuchi02}.  As these particles drift inwards, their size distribution is 
continuously evolving as they encounter higher local densities that are 
conducive to growth \citep{birnstiel10}.  The collective effect is to shrink the
radial extent of the dust mass distribution, as the gas disk continues to
viscously spread to larger radii \citep{birnstiel12a}.  Since these processes 
operate on dynamical timescales, this phenomenon should be particularly 
pronounced for the V4046 Sgr disk due to the binarity (increased $M_*$) and 
advanced age of its stellar hosts.  Resolved observations taken across the 
(sub-)millimeter and radio bands may clarify whether the radial grain size 
distribution is consistent with this drift and particle growth scenario 
\citep{guilloteau11,perez12}.

In order to reproduce the observed morphology of the mm continuum emission, we 
required a model that has its large dust grain population distributed in a 
narrow ring.  This is suggestive of a physical mechanism acting to both
concentrate the particles radially and prevent them from moving further into 
the inner disk.  Particle growth alone is insufficient to produce mm-wave 
cavities \citep{birnstiel12b}, but this kind of ring-like dust concentration is 
a natural consequence of particle evolution models that also employ feedback 
from a large-scale maximum in the radial pressure gradient of the gas 
\citep{pinilla12a}.  The dynamics of the dust particles are strongly dictated by
the surrounding gas and while negative radial pressure gradients induce inward 
radial drift, a {\it positive} pressure gradient may stop the dust as it streams
towards the star.  This pressure maximum forms a ``pocket'' that collects dust 
and, if sufficiently broad and high-amplitude, can significantly alter the 
observed continuum emission \citep{pinilla12b}.

While many physical phenomena may generate these particle traps 
\citep[e.g.][]{barge95,klahr97,alexander07,johansen09}, we focus on an origin in
the density maximum at a gap edge produced by dynamical interactions between a 
low-mass companion embedded in the gas disk \citep{lin79,goldreich80,crida06}.  
Simulations by \citet{pinilla12a} produced a ring of large grains in a broad 
surface density bump exterior to the radius of such a gap.  Furthermore, their 
models of the mutual evolution of the gas and dust predicted a strong radial 
variation in the dust-to-gas mass ratio (see their Figure 6) with heavily 
depleted dust densities in the outer disk ($\zeta \sim 10^{-3}$ -- $10^{-4}$) 
and a large concentration within the trap ($\zeta \sim 1$).  Both the 
ring-shaped surface density profile and radially decreasing dust-to-gas ratio 
produced by their calculation were necessary features of the model we derived 
for the V4046 Sgr disk.  

\begin{figure}[t!]
\epsscale{0.6}
\plotone{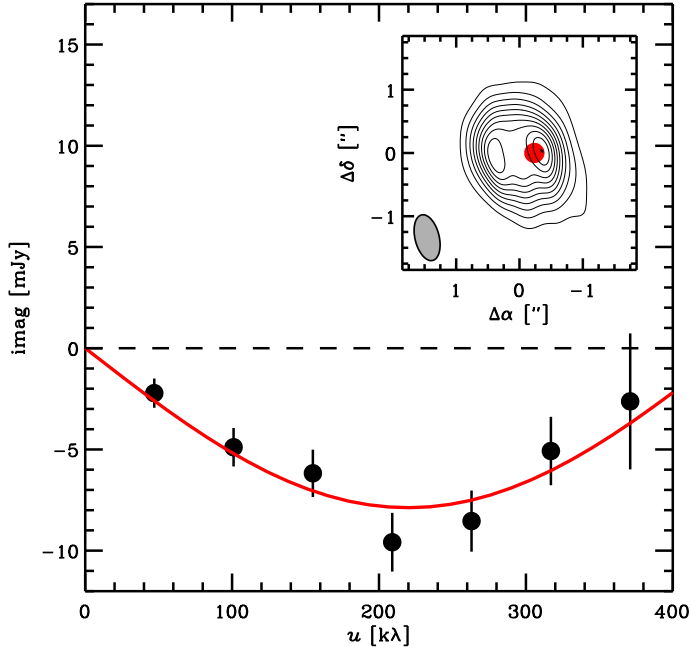}
\figcaption{ The complex visibilities of the 1.3\,mm emission binned along the 
$v$ axis ({\it black points}) with a model 8\,mJy point source ({\it red}) 
offset by $\Delta \alpha = -0\farcs23$ from the disk center ({\it red curve}).  
The inset shows the continuum image with 5\,$\sigma$ contour levels and the 
model point source ({\it red dot}).
\label{fig:asym}}
\end{figure}

Recent models that account for the expected substructure in these ring-like 
pressure traps suggest that a modest variation in the gas density 
(peak-to-valley amplitude ratio of $\gtrsim$1.5) can produce strong azimuthal 
asymmetries in the dust distribution \citep{regaly12,birnstiel13}.  This 
phenomenon may be sufficient to qualitatively explain some observations of the 
``lopsided'' emission from some transition disks \citep[e.g.][]{brown09,
mayama12,casassus13,vandermarel13}.  As noted in \S\ref{sec:results}, the western peak of the 
V4046 Sgr dust ring appears slightly brighter than the eastern peak by 
$\sim 5$\,mJy\,beam$^{-1}$.  We find that this subtle asymmetry is apparent in 
the continuum visibilities as well, and therefore is not likely to be an 
artifact of the imaging/deconvolution process.  Figure \ref{fig:asym} shows a 
cut of the imaginary component of the complex visibilities binned along the 
$v$-axis in the Fourier plane (corresponding to the E-W direction).  For a 
symmetric morphology, the imaginary visibilities should be zero: however, we 
find a statistically significant deviation in a sinusoid pattern.  The 
asymmetry can be reproduced reasonably well with a faint point source 
($\sim$8\,mJy) located $\sim$0\farcs23 to the west of the disk center.  However,
verification and characterization of this low-level apparent asymmetry beyond 
such a simple model will require observations with higher resolution and 
sensitivity.

The ring-shaped dust disk and tentative asymmetry seem to be well-matched to 
predictions for a large pressure trap generated by a companion interacting with 
the gas disk.  However, the central spectroscopic binary itself is too compact 
and circular to have dynamically truncated the disk at such a large radius: this
scenario requires (at least) a third body.  One constraint on the properties of 
a putative companion comes from the dynamical mass estimate by 
\citet{rosenfeld12}.  Since this method utilizes the Keplerian rotation pattern
of the disk, it is sensitive to the cumulative mass at the disk center: the 
spectroscopic binary plus any other companion(s).  If the central binary and 
disk are co-planar, \citet{rosenfeld12} showed that the disk-based dynamical 
mass estimate was consistent with that inferred from spectroscopic monitoring of
the stellar orbits \citep{stempels13}.  Therefore, any additional companion must
have a low mass: the 3\,$\sigma$ uncertainties on the dynamical mass estimate of
\citet{rosenfeld12} require that $M_{\rm{comp}} \le 0.3$\,M$_\odot$.  A second,
complementary constraint can be made from the contribution of any companion to 
the near-infrared spectrum.  The composite stellar photosphere model adopted 
here accounts for 99, 95, and 85\%\ of the observed emission in the $K$, 
$L^{\prime}$ and $M$ bands, respectively \citep{jensen97,hutchinson90,
skrutskie06}.  If we attribute the remaining emission to a co-eval low-mass 
companion, rather than the tenuous inner dust ring described in \S 4, pre-main 
sequence models \citep[e.g.][]{baraffe98} suggest that the spectrum requires
$M_{\rm comp} \le 0.2$\,M$_\odot$.

\begin{figure}[t!]
\epsscale{0.6}
\plotone{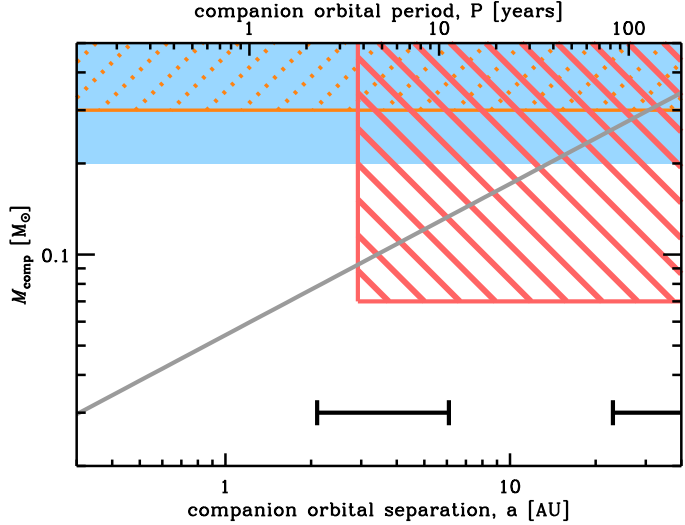}
\figcaption{Limits on a companion mass with regions ruled out by the dynamical 
mass estimate \citep[{\it orange dots};][]{rosenfeld12}, infrared luminosities 
\citep[{\it solid blue};][]{jensen97,hutchinson90,skrutskie06}, and infrared 
aperture-masking observations ({\it thick rose lines}; A.~L.~Kraus, private 
communication).  The lower mass limit consistent with the reported velocity 
shifts is also shown \citep[{\it gray line};][]{quast00,donati11,rodriguez10}. 
The position and 2\,$\sigma$ widths of the main and inner dust rings are marked 
by floating error bars.
\label{fig:companion}}
\end{figure}

A third constraint on a potential companion was suggested by \citet{donati11}, 
based on their claim of a shift in the systemic velocity of the spectroscopic 
binary.  They measured $V_{\rm sys}=-5.7\pm0.2$\,km s$^{-1}$ (in the 
heliocentric frame) at a mean epoch of 2009.7, which is significantly different
than the previous measurements of the binary \citep[$-6.94\pm0.01$\,km 
s$^{-1}$ at epoch 1985.5;][]{quast00} and the gas disk 
\citep[$-6.26\pm.05$\,km s$^{-1}$;][]{rodriguez10}.  Assuming the total mass 
determined by \citet{rosenfeld12} and co-planar orbits, we can derive a lower 
limit on the companion mass as a function of semi-major axis (or orbital period)
from these apparent velocity shifts: $M_{\rm comp} \gtrsim 0.05$\,M$_\odot 
(a/{\rm 1AU})^{1/2}$, where $a$ is the semi-major axis.  It should be 
noted that these apparent velocity shifts are likely consistent with one another
when considering the typical systematic uncertainties in the velocity
calibration between different instruments.  A final, more direct, constraint on
the companion mass can be made from infrared aperture-masking observations 
(A.~L.~Kraus; private communication). No companion is detected in those data, 
with a $K$-band contrast limit of 6-7 magnitudes outside of 0\farcs04 from the 
central binary.  Assuming a co-eval companion and the \citet{baraffe98} models,
these data constrain $M_{\rm comp} \le 0.07$\,M$_\odot$ for $a>2.9$\,AU.  
Figure \ref{fig:companion} summarizes these four constraints on $M_{\rm comp}$ 
and $a$.

Given the results in Figure \ref{fig:companion}, it is clear that the mass of 
any companion must be low, making a low-mass brown dwarf or massive giant planet
likely candidates.  The position of this hypothetical planet from the observed 
ring is highly uncertain and with no conclusive evidence for reduced densities 
in the cavity, our CO observations cannot help constrain $M_{\rm comp}$ or $a$. 
However, sensitive observations with high angular resolution may resolve 
structure in the inner gas disk \citep{casassus13} or signatures of clearing in 
the high-velocity wings \citep{dutrey08}.  An alternative way of
locating a companion is by looking at the small grains that filter past it 
and using this flow of small grains entrained in the gas to quantify the mass of
the companion \citep{rice06,zhu12}.  While the distribution of $\mu$m-sized dust
particles in our model are closer to the central binary than the inner working
angle of the SEEDS survey \citep[0\farcs1-0\farcs15]{tamura09},
the V4046 Sgr disk is a natural candidate to test this scenario with scattered 
light observations \citep[e.g.][]{dong12}.

A compelling potential alternative mechanism for shaping the distinctive 
structure of the V4046 Sgr disk is photoevaporation \citep{hollenbach94,
clarke01,alexander06a,alexander06b,owen12} which can also create large pressure 
traps \citep{alexander07}.  Strong photoionizing radiation from the central 
stars drive a wind off the disk surface, and consequently can open a gap in the 
disk on $\sim$AU scales.  V4046 Sgr has a large UV excess \citep{delareza86,
hutchinson90,huenemoerder07} and large X-ray luminosity 
\citep[$L_X \sim (7-10)\times 10^{29}$\,erg s$^{-1}$, based on archival XMM 
data; see also][]{gunther06,argiroffi12}.  According to recent models
\citep{gorti09a,gorti09b,owen10,owen11}, this high-energy emission should drive 
a photoevaporative wind.  Indeed, the presence of such a wind has been 
inferred via analysis of mid-infrared [Ne II] emission from V4046 Sgr 
\citep{sacco12}.  However, given the suite of models presented by 
\citet{owen11} and the present accretion rate of V4046 Sgr \cite[$\log\dot{M}
= -9.3\pm0.3$\,M$_\odot$ yr$^{-1}$;][]{donati11,curran11}, the X-ray luminosity 
is too low to explain the large cavity size inferred here.  However, X-ray 
photoevaporation (and perhaps even EUV/FUV photoevaporation) might find a 
natural role in explaining the depletion of material inside the innermost dust 
ring ($r \lesssim 3$\,AU inferred from the SED, see \S\ref{sec:model1}).

Perhaps the greatest mystery surrounding the V4046 Sgr disk is its existence: 
most theoretical and observational work suggests that disk dispersal processes 
should substantially diminish the observational signatures studied here well 
within the lifetime of this system and the few others like it \citep{clarke01,
alexander06b,takeuchi05,currie07,currie09}.  The modeling analysis 
conducted here unfortunately is unable to directly answer the question of why 
this gas-rich disk persists even at an age of 10-20\,Myr.  However, the inferred
disk structure might be providing some clues.  It might be possible that the 
cavity was formed early in the disk lifetime, stifling the viscous evolution 
process by severely limiting the gas accretion flow onto the central stars, and
thereby slowing the natural evolution of the disk material.  Alternatively, 
\citet{alexander12} suggested that the tidal torque induced by a close binary 
on its disk may inhibit accretion and increase the lifetime of the circumbinary 
disk compared to that of a disk around a single star. Such scenarios are 
speculative: but it is interesting to note that most (if not all) of the known 
long-lived, gas-rich disks have evidence for low-density cavities at their 
centers.

Overall, the signatures of evolution in the transition disk hosted by V4046 Sgr
are remarkable, suggesting that this target merits a focused effort moving 
forward with ALMA.  Its close proximity and large physical size enables high 
resolution and high sensitivity observations.  Furthermore, the stellar 
binarity aids the observations: it is too tight to affect the disk structure, 
but the high stellar mass means that the widths of gas disk emission lines is 
large and so spectral line observations can have good velocity resolution 
without sacrificing sensitivity.  Observations are further aided by the fact 
that it is an isolated system unaffected by cloud contamination.  With all of 
this in mind, V4046 Sgr promises to be an excellent laboratory for studying disk
evolution and planet formation: its old age means that evolutionary features are
well-advanced, and its rich gas disk provides an excellent target for both 
detection efforts and spatially resolved analysis.  Lastly, the X-ray luminosity
makes V4046 Sgr a natural target for studying photoevaporation and its effects 
on disk evolution.  

\section{Summary}

We have studied the structure of the V4046 Sgr circumbinary disk using high 
resolution, high sensitivity mm-wave observations from the SMA.  By modeling the
SED, spatially resolved 1.3\,mm dust continuum, and line emission of three CO 
isotopes, we have identified this disk as an exemplar evolved protoplanetary 
disk.  The key conclusions of our analysis are:

\begin{enumerate}
\item The CO gas disk is significantly more extended than the compact 
distribution of large grains.  In order to reproduce this feature we decouple 
the dust and gas surface density profiles, reducing the dust-to-gas ratio in the
outskirts of our disk model by a factor of 10.  We interpret this phenomenon as 
the result of the simultaneous growth and inward radial drift of mm/cm-sized 
particles in a viscously evolving gas disk \citep{birnstiel12a}.

\item The V4046 Sgr disk features a large central cavity ($r=29$\,AU) depleted 
of mm/cm-sized dust grains.  These grains are concentrated in a narrow ring 
($\mu=37$\,AU, FWHM=$16$\,AU).  There is no conclusive evidence from the CO 
line emission whether there is an analogous depletion in the gas disk. This 
distinctive morphology is consistent with simulations that feature large 
pressure traps generated by the dynamical interaction of a massive companion 
with a gas disk and the evolution of solid material \citep{pinilla12a}.  

\item The cavity contains a significant population of $\mu$m-sized grains.  
We infer that this distribution of small dust particles has a smaller central
clearing that may be consistent with X-ray photoevaporation \citep{owen11}, but 
is too large to be truncated by the central binary \citep{artymowicz94}.  This 
population might originate from small grains entrained in the gas flowing 
past the pressure barrier that traps the mm/cm-sized grains in the large ring
\citep{rice06}.

\end{enumerate}

\acknowledgments We are grateful to Adam Kraus for sharing his $K$-band limit on
a companion, Andrea Isella for teaching K.R. about asymmetries, and Moritz 
G{\"u}nther for helping us with the X-ray luminosities.  We are also indebted to
Kees Dullemond, Guillermo Torres, and Matt Payne for insightful discussions.  
We acknowledge support from NASA Origins of Solar Systems grant No. NNX11AK63.  
J.K.'s research on V4046 Sgr and other nearby protoplanetary disks is supported 
in part by National Science Foundation grant AST-1108950 to RIT.  The 
Submillimeter Array is a joint project between the Smithsonian Astrophysical 
Observatory and the Academia Sinica Institute of Astronomy and Astrophysics and 
is funded by the Smithsonian Institution and the Academia Sinica. This work is 
based [in part] on archival data obtained with the Spitzer Space Telescope, 
which is operated by the Jet Propulsion Laboratory, California Institute of 
Technology under a contract with NASA. Support for this work was provided by 
NASA. 

\clearpage

\begin{deluxetable}{lcccccc}
\tablecolumns{7}
\tablewidth{0pc}
\tablecaption{Summary of Observations\label{tab:obs}}
\tablehead{
\colhead{UT date} & \colhead{configuration} & \colhead{antennas} & \colhead{baselines} & \colhead{$t_{\rm on}$} & \colhead{$\Delta \nu_{\rm cont}$} & \colhead{$\tau_{\rm atm}$} \\
\colhead{} & \colhead{} & \colhead{} & \colhead{[k$\lambda$]} & \colhead{[min]} & \colhead{[GHz]} & \colhead{(225\,GHz)}
}
\startdata
2009 Feb 23 & extended      & 8 & 17-132 & 110 & 3.4 & 0.05 \\
2009 Apr 25 & compact-N     & 6 & 6-84   & 220 & 3.4 & 0.06 \\
2011 Mar  8 & sub-compact   & 7 & 4-50   & 90  & 7.4 & 0.05 \\
2011 Sep  4 & very extended & 8 & 19-390 & 180 & 7.4 & 0.04
\enddata
\end{deluxetable}


\clearpage


\begin{thebibliography}{}
\bibitem[Aikawa \& Herbst(1999)]{aikawa99} Aikawa, Y., \& Herbst, E. 1999, \aap, 351, 233
\bibitem[Alexander et al.(2004)]{alexander04} Alexander, R.~D., 
Clarke, C.~J., \& Pringle, J.~E.\ 2004, \mnras, 354, 71 
\bibitem[Alexander et al.(2006a)]{alexander06a} Alexander, R.~D., 
Clarke, C.~J., \& Pringle, J.~E.\ 2006, \mnras, 369, 216
\bibitem[Alexander et al.(2006b)]{alexander06b} Alexander, R.~D., 
Clarke, C.~J., \& Pringle, J.~E.\ 2006, \mnras, 369, 229 
\bibitem[Alexander \& Armitage(2007)]{alexander07} Alexander, R.~D., \& Armitage, P.~J.\ 2007, \mnras, 375, 500 
\bibitem[Alexander \& Armitage(2009)]{alexander09} Alexander, R. D., \& Armitage, P. J. 2009, \apj, 704, 989
\bibitem[Alexander(2012)]{alexander12} Alexander, R.\ 2012, \apjl, 757, L29 
\bibitem[Andrews et al.(2009)]{andrews09} Andrews, S. M., Wilner, D. J., Hughes, A. M., Qi, C., \& Dullemond, C. P. 2009, \apj, 700, 1502
\bibitem[Andrews et al.(2010)]{andrews10} Andrews, S. M., Wilner, D. J., Hughes, A. M., Qi, C., \& Dullemond, C. P. 2010, \apj, 723, 1241
\bibitem[Andrews et al.(2011a)]{andrews11} Andrews, S. M., et al. 2011a, \apj, 732, 42 (2011)
\bibitem[Andrews et al.(2011b)]{andrews11b} Andrews, S. M., Rosenfeld, K. A., Wilner, D. J., \& Bremer, M. 2011b, \apj, 742, L5 (2011)
\bibitem[Andrews et al.(2012)]{andrews12} Andrews, S. M., et al. 2012, \apj, 744, 162
\bibitem[Argiroffi et al.(2012)]{argiroffi12} Argiroffi, C., Maggio, A., Montmerle, T., et al.\ 2012, \apj, 752, 100
\bibitem[Artymowicz \& Lubow(1994)]{artymowicz94} Artymowicz, P., \& Lubow, S. H. 1994, \apj, 421, 651
\bibitem[Baraffe et al.(1998)]{baraffe98} Baraffe, I., Chabrier, G., Allard, F., \& Hauschildt, P.~H.\ 1998, \aap, 337, 403
\bibitem[Barge \& Sommeria(1995)]{barge95} Barge, P., \& Sommeria, J.\ 1995, \aap, 295, L1 
\bibitem[Beichman et al.(1988)]{beichmann88} Beichmann, C. A., Neugebauer, G., Habing, H. J., Clegg, P. E., \& Chester, T. J. 1988, in Infrared Astronomical Satellie Catalogs and Atlases.  Volume 1: Explanatory Supplement
\bibitem[Birnstiel et al.(2010)]{birnstiel10} Birnstiel, T., Dullemond, C.~P., \& Brauer, F.\ 2010, \aap, 513, A79 
\bibitem[Birnstiel et al.(2012a)]{birnstiel12a} Birnstiel, T., Klahr, H., \& Ercolano, B. 2012a, \aap, 539, 148
i\bibitem[Birnstiel et al.(2012b)]{birnstiel12b} Birnstiel, T., Andrews, S.~M., \& Ercolano, B.\ 2012b, \aap, 544, A79
\bibitem[Birnstiel et al.(2013)]{birnstiel13} Birnstiel, T., Dullemond, C.~P., \& Pinilla, P.\ 2013, \aap, 550, L8 
\bibitem[Brauer et al.(2008)]{brauer08a} Brauer, F., Dullemond, C.~P., \& Henning, T.\ 2008, \aap, 480, 859
\bibitem[Brinch \& Hogerheijde(2010)]{brinch10} Brinch, C., \& Hogerheijde, M. R. 2010, \aap, 523, 25
\bibitem[Brown et al.(2009)]{brown09} Brown, J.~M., Blake, G.~A., Qi, C., et al.\ 2009, \apj, 704, 496 
\bibitem[Brown et al.(2012)]{brown12} Brown, J. M., Rosenfeld, K. A., Andrews, S. M., Wilner, D. J., \& van Dishoeck, E. F. 2012, \apj, 758, L30
\bibitem[Byrne(1986)]{byrne86} Byrne, P. B. 1986, Ir. Astron. J., 17, 294
\bibitem[Calvet et al.(2002)]{calvet02} Calvet, N., D'Alessio, P., Hartmann, L., Wilner, D., Walsh, A., \& Sitko, M. 2002, \apj, 568, 1008
\bibitem[Casassus et al.(2013)]{casassus13} Casassus, S., van der 
Plas, G., M, S.~P., et al.\ 2013, \nat, 493, 191
\bibitem[Clarke et al.(2001)]{clarke01} Clarke, C. J., Gendrin, A., \& Sotomayor, M. 2001, \mnras, 328, 485
\bibitem[Crida et al.(2006)]{crida06} Crida, A., Morbidelli, 
A., \& Masset, F.\ 2006, \icarus, 181, 587 
\bibitem[Curran et al.(2011)]{curran11} Curran, R.~L., Argiroffi, C., Sacco, G.~G., et al.\ 2011, \aap, 526, A104 
\bibitem[Currie et al.(2007)]{currie07} Currie, T., Balog, Z., 
Kenyon, S.~J., et al.\ 2007, \apj, 659, 599 
\bibitem[Currie et al.(2009)]{currie09} Currie, T., Lada, C.~J., Plavchan, P., et al.\ 2009, \apj, 698, 1 
\bibitem[D'Alessio et al.(2001)]{dalessio01} D'Alessio, P., Calvet, N., \& Hartmann, L.\ 2001, \apj, 553, 321 
\bibitem[Dartois et al.(2003)]{dartois03} Dartois, E., Dutrey, A., \& Guilloteau, S. 2003, \aap, 399, 773
\bibitem[Dong et al.(2012)]{dong12} Dong, R., Rafikov, R., Zhu, Z., et al.\ 2012, \apj, 750, 161
\bibitem[de la Reza et al.(1986)]{delareza86} de la Reza, R., Quast, G., Torres, C. A. O., Mayor, M., Meylan, G., \& Llorente de Andres, F. 1986, ESASP, 263, 107
\bibitem[Donati et al.(2011)]{donati11} Donati, J.-F., Gregory, S.~G., Montmerle, T., et al.\ 2011, \mnras, 417, 1747
\bibitem[Dutrey et al.(2008)]{dutrey08} Dutrey, A., Guilloteau, S., Pi{\'e}tu, V., et al.\ 2008, \aap, 490, L15
\bibitem[Goldreich \& Tremaine(1980)]{goldreich80} Goldreich, P., \& Tremaine, S.\ 1980, \apj, 241, 425 
\bibitem[Gorti \& Hollenbach(2009)]{gorti09a} Gorti, U., \& Hollenbach, D.\ 2009, \apj, 690, 1539 
\bibitem[Gorti et al.(2009)]{gorti09b} Gorti, U., Dullemond, 
C.~P., \& Hollenbach, D.\ 2009, \apj, 705, 1237
\bibitem[Guilloteau et al.(2011)]{guilloteau11} Guilloteau, S., Dutrey, A., Pi{\'e}tu, V., \& Boehler, Y. 2011, \aap, 529, 105
\bibitem[G{\"u}nther et al.(2006)]{gunther06} G{\"u}nther, H.~M., Liefke, C., Schmitt, J.~H.~M.~M., Robrade, J., \& Ness, J.-U.\ 2006, \aap, 459, L29
\bibitem[Harris et al.(2012)]{harris12} Harris, R. J., Andrews, S. M., Wilner, D. J., \& Kraus, A. L. 2012, \apj, 751, 115
\bibitem[Hartmann et al.(1998)]{hartmann98} Hartmann, L., Calvet, N., Gullbring, E., \& D'Alessio, P.\ 1998, \apj, 495, 385
\bibitem[Henize(1976)]{henize76} Henize, K. G. 1976, \apjs, 30, 491
\bibitem[Herbig \& Bell(1988)]{hbc88} Herbig, G. H., \& Bell, K. R. 1988, Third Catalog of Emission-Line Stars of the Orion Population, Lick Observatory Bulletin \#1111 (Santa Cruz, CA)
\bibitem[Ho et al.(2004)]{ho04} Ho, P. T. P., Moran, J. M., \& Lo, K. Y. 2004, \apj, 616, L1
\bibitem[Hollenbach et al.(1994)]{hollenbach94} Hollenbach, D., Johnstone, D., Lizano, S., \& Shu, F.\ 1994, \apj, 428, 654
\bibitem[Hubickyj et al.(2005)]{hubickyj05} Hubickyj, O., Bodenheimer, P., \& Lissauer, J. J. 2005, Icarus, 179, 415
\bibitem[Huenemoerder et al.(2007)]{huenemoerder07} Huenemoerder, D.~P., Kastner, J.~H., Testa, P., Schulz, N.~S., \& Weintraub, D.~A.\ 2007, \apj, 671, 592 
\bibitem[Hutchinson et al.(1990)]{hutchinson90} Hutchinson, M. G., Evans, A., Winkler, H., \& Spencer Jones, J. 1990, \aap, 234, 230
\bibitem[Isella et al.(2009)]{isella09} Isella, A., Carpenter, J. M., \& Sargent, A. I. 2009, \apj, 701, 260
\bibitem[Isella et al.(2010)]{isella10} Isella, A., Natta, A., Wilner, D., Carpenter, J. M., \& Testi, L. 2010, \apj, 725, 1735
\bibitem[Isella et al.(2012)]{isella12} Isella, A., P{\'e}rez, L. M., \& Carpenter, J. M. 2012, \apj, 747, 136
\bibitem[Jensen et al.(1996)]{jensen96} Jensen, E. L. N., Mathieu, R. D., \& Fuller, G. A. 1996
\bibitem[Jensen \& Mathieu(1997)]{jensen97} Jensen, E. L. N., \& Mathieu, R. D. 1997, \aj, 114, 301
\bibitem[Johansen et al.(2009)]{johansen09} Johansen, A., Youdin, A., \& Klahr, H.\ 2009, \apj, 697, 1269 
\bibitem[Johnson(1986)]{johnson86} Johnson, H. M. 1986, \apj, 300, 401
\bibitem[Kastner et al.(1997)]{kastner97} Kastner, J.~H., Zuckerman, B., Weintraub, D.~A., \& Forveille, T.\ 1997, Science, 277, 67 
\bibitem[Kastner et al.(2008)]{kastner08} Kastner, J. H., Zuckerman, B., Hily-Blant, P., \& Forveille, T. 2008, \aap, 492, 469
\bibitem[Kastner et al.(2010)]{kastner10} Kastner, J.~H., Hily-Blant, P., Sacco, G.~G., Forveille, T., \& Zuckerman, B.\ 2010, \apjl, 723, L248 
\bibitem[Kastner et al.(2011)]{kastner11} Kastner, J.~H., Sacco,
G.~G., Montez, R., et al.\ 2011, \apjl, 740, L17
\bibitem[Klahr \& Henning(1997)]{klahr97} Klahr, H.~H., \& Henning, T.\ 1997, \icarus, 128, 213 
\bibitem[Lejeune et al.(1997)]{lejeune97} Lejeune, T., Cuisinier, F., \& Buser, R.\ 1997, \aaps, 125, 229
\bibitem[Lin \& Papaloizou(1979)]{lin79} Lin, D.~N.~C., \& Papaloizou, J.\ 1979, \mnras, 188, 191 
\bibitem[Lynden-Bell \& Pringle(1974)]{lynden-bell74} Lynden-Bell, D., \& Pringle, J. E. 1974, \mnras, 168, 603
\bibitem[Mayama et al.(2012)]{mayama12} Mayama, S., Hashimoto, J., Muto, T., et al.\ 2012, \apjl, 760, L26 
\bibitem[Merrill \& Burwell(1950)]{merrill50} Merrill, P.~W., \& Burwell, C.~G.\ 1950, \apj, 112, 72 
\bibitem[Nataf et al.(2010)]{nataf10} Nataf, D. M., Stanek, K. Z., \& Bakos, G. {\'A}. 2010, Acta Astronomica, 60, 261
\bibitem[Owen et al.(2010)]{owen10} Owen, J.~E., Ercolano, B., Clarke, C.~J., \& Alexander, R.~D.\ 2010, \mnras, 401, 1415 
\bibitem[Owen et al.(2011)]{owen11} Owen, J.~E., Ercolano, B., \& Clarke, C.~J.\ 2011, \mnras, 412, 13
\bibitem[Owen et al.(2012)]{owen12} Owen, J.~E., Clarke, C.~J., \& Ercolano, B.\ 2012, \mnras, 422, 1880
\bibitem[{\"{O}}berg et al.(2011)]{oberg11} {\"{O}}berg, K. I., et al. 2011, \apj, 734, 98
\bibitem[Pani{\'c} et al.(2008)]{panic08} Pani{\'c}, O., Hogerheijde, M.~R., Wilner, D., \& Qi, C.\ 2008, \aap, 491, 219 
\bibitem[Pani{\'c} et al.(2009)]{panic09} Pani{\'c}, O., Hogerheijde, M. R., Wilner, D., \& Qi, C. 2009, \aap, 501, 269
\bibitem[P{\'e}rez et al.(2012)]{perez12} P{\'e}rez, L. M., et al. 2012, \apj, 760, L17
\bibitem[Pinilla et al.(2012a)]{pinilla12a} Pinilla, P., Benisty, M., \& Birnstiel, T.\ 2012, \aap, 545, A81 
\bibitem[Pinilla et al.(2012b)]{pinilla12b} Pinilla, P., Birnstiel, T., Ricci, L., et al.\ 2012b, \aap, 538, A114
\bibitem[Pollack et al.(1994)]{pollack94} Pollack, J. B., Hollenbach, D., Beckwith, S., Simonelli, D. P., Roush, T., \& Fong, W. 1994, \apj, 421, 615
\bibitem[Pollack et al.(1996)]{pollack96} Pollack, J. B., Hubickyj, O., Bodenheimer, P., Lissauer, J. J., Podolak, M., \& Greenzweig, Y. 1996, Icarus, 124, 62
\bibitem[Qi et al.(2008)]{qi08} Qi, C., Wilner, D. J., Aikawa, Y., Blake, G. A., \& Hogerheijde, M. R. 2008, \apj, 681, 1396
\bibitem[Qi et al.(2011)]{qi11} Qi, C., et al. 2011, \apj, 740, 84
\bibitem[Quast et al.(2000)]{quast00} Quast, G. R., Torres, C. A. O., de La Reza, R., da Silva, L., \& Mayor, M. 2000, IAU Symposium, 200, 28P
\bibitem[Reg{\'a}ly et al.(2012)]{regaly12} Reg{\'a}ly, Z., Juh{\'a}sz, A., S{\'a}ndor, Z., \& Dullemond, C.~P.\ 2012, \mnras, 419, 1701 
\bibitem[Rice et al.(2006)]{rice06} Rice, W.~K.~M., Armitage, 
P.~J., Wood, K., \& Lodato, G.\ 2006, \mnras, 373, 1619
\bibitem[Rodriguez et al.(2010)]{rodriguez10} Rodriguez, D. R., Kastner, J. H., Wilner, D., \& Qi, C. 2010, \apj, 720, 1684
\bibitem[Rosenfeld et al.(2012)]{rosenfeld12} Rosenfeld, K. A., Andrews, S. M., Wilner, D. J., \& Stempels, H. C. 2012a, \apj, 759, 119
\bibitem[Sacco et al.(2012)]{sacco12} Sacco, G.~G., Flaccomio, E., Pascucci, I., et al.\ 2012, \apj, 747, 142 
\bibitem[Sacco et al.(2013)]{sacco13} Sacco. G.~G., et al.\ 2013, in preparation
\bibitem[Sch{\"o}ier et al.(2005)]{schoier05} Sch{\"o}ier, F. L., van der Tak, F. F. S., van Dishoeck, E. F., \& Black, J. H. 2005, \aap, 432, 369
\bibitem[Skrutskie et al.(2006)]{skrutskie06} Skrutskie, M. F., et al. 2006, \aj, 131, 1163
\bibitem[Stempels \& Gahm(2004)]{stempels04} Stempels, H. C., \& Gahm, G. F. 2004, \aap, 421, 1159
\bibitem[Stempels(2013)]{stempels13} Stempels, H. C. 2013, in preparation
\bibitem[Takeuchi \& Lin(2002)]{takeuchi02} Takeuchi, T., \& Lin, D.~N.~C.\ 2002, \apj, 581, 1344 
\bibitem[Takeuchi et al.(2005)]{takeuchi05} Takeuchi, T., Clarke, 
C.~J., \& Lin, D.~N.~C.\ 2005, \apj, 627, 286 
\bibitem[Tamura(2009)]{tamura09} Tamura, M.\ 2009, American Institute of Physics Conference Series, 1158, 11 
\bibitem[Torres et al.(2006)]{torres06} Torres, C. A. O., Quast, G. R., da Silva, L., de la Reza, R., Melo, C. H. F., \& Sterzik, M. 2006, \aap, 460, 695
\bibitem[Torres et al.(2008)]{torres08} Torres, C.~A.~O., Quast,
G.~R., Melo, C.~H.~F.,
\& Sterzik, M.~F.\ 2008, Handbook of Star Forming Regions, Volume II, 757
\bibitem[van der Marel et al.(2013)]{vandermarel13} van der Marel, 
N., van Dishoeck, E.~F., Bruderer, S., et al.\ 2013, Science, 340, 1199 
\bibitem[Weaver \& Jones(1992)]{weaver92} Weaver, W. B., \& Jones, G. 1992, \apjs, 78, 239
\bibitem[Weidenschilling(1977)]{weidenschilling77} Weidenschilling, 
S.~J.\ 1977, \mnras, 180, 57
\bibitem[Weidenschilling(1980)]{weidenschilling80} Weidenschilling, 
S.~J.\ 1980, \icarus, 44, 172 
\bibitem[Weintraub(1990)]{weintraub90} Weintraub, D. A. 1990, \apjs, 74, 575
\bibitem[Wilson(1999)]{wilson99} Wilson, T. L. 1999, Reports on Progress in Physics, 62, 143
\bibitem[Zacharias et al.(2010)]{zacharias10} Zacharias, N., et al. 2010, \aj, 139, 2184
\bibitem[Zhu et al.(2012)]{zhu12} Zhu, Z., Nelson, R.~P., 
Dong, R., Espaillat, C., \& Hartmann, L.\ 2012, \apj, 755, 6
\end{thebibliography}
\end{document}